\pdfoutput=1

\documentclass[pdflatex,sn-nature]{sn-jnl}
\usepackage{graphicx}%
\usepackage{multirow}%
\usepackage{amsmath,amssymb,amsfonts}%
\usepackage{amsthm}%
\usepackage{mathrsfs}%
\usepackage[title]{appendix}%
\usepackage{xcolor}%
\usepackage{textcomp}%
\usepackage{manyfoot}%
\usepackage{booktabs}%
\usepackage{algorithm}%
\usepackage{algorithmicx}%
\usepackage{algpseudocode}%
\usepackage{listings}%

\newcommand\ho{\ifmmode {\rm HI} \else H{\small I} \fi}
\newcommand\hh{\ifmmode {\rm H_2} \else H$_2$ \fi}
\def\no{\ifmmode {N_{\rm HI}} \else $N_{\rm HI}$ \fi}
\def\nt{\ifmmode {N_{\rm H_2}} \else $N_{\rm HI}$ \fi}
\def\msun{\ifmmode {\rm M_{\odot}}\else $\rm M_{\odot}$\fi}
\def\mpc{\ifmmode {\rm M_{\odot} \ pc^{-2}} \else $\rm M_{\odot} \ pc^{-2}$ \fi}
\def\tra{\ifmmode  \text{HI-to-H}_2\else H{\small I}-to-H$_2$ \fi}
\def\aG{\ifmmode {\alpha G}\else $\alpha G$ \fi}
\def\iuv{\ifmmode {I_{\rm UV}}\else $I_{\rm UV}$ \fi}

\def\sg{\ifmmode \sigma_{g} \else $\sigma_{g}$ \fi}
\newcommand\hd{\ifmmode \textrm{HI-dust} \else H{\small I}-dust \fi}
\DeclareMathAlphabet{\pazocal}{OMS}{zplm}{m}{n}
\newcommand\ms{\ifmmode \pazocal{M}_s \else $\pazocal{M}_s$ \fi}
\newcommand\Nm{\ifmmode  N_{\rm M}  \else $N_{\rm M}$ \fi}

\renewcommand\refname{References and Notes}

\begin{document}

\title{A Nearby Dark Molecular Cloud in the Local Bubble Revealed via H$_2$ Fluorescence}

\author*[1,2]{\fnm{Blakesley } \sur{Burkhart}}\email{bburkhart@flatironinstitute.org}
\equalcont{These authors contributed equally to this work.}
\author[2,3]{\fnm{Thavisha} \sur{E. Dharmawardena}}\email{tdharmawardena@flatironinstitute.org}
\equalcont{These authors contributed equally to this work.}

\author[4]{\fnm{Shmuel} \sur{Bialy}}\email{sbialy@technion.ac.il}
\author[5]{\fnm{Thomas} \sur{J. Haworth}}\email{t.haworth@qmul.ac.uk}
\author[6]{\fnm{Fernando} \sur{Cruz Aguirre}}\email{edwinfernando-cruzaguirre@uiowa.edu}
\author[7]{\fnm{Young-Soo} \sur{Jo}}\email{stspeak@kasi.re.kr}
\author[8]{\fnm{B-G} \sur{Andersson}}\email{astrobgandersson@gmail.com}
\author[9]{\fnm{Haeun} \sur{Chung}}\email{haeunchung@arizona.edu}
\author[10]{\fnm{Jerry} \sur{Edelstein}}\email{jerrye@berkeley.edu}
\author[11]{\fnm{Isabelle} \sur{Grenier}}\email{isabelle.grenier@cea.fr}
\author[9]{\fnm{Erika} \sur{T. Hamden}}\email{hamden@arizona.edu}
\author[7]{\fnm{Wonyong} \sur{Han}}\email{whan@kasi.re.kr}
\author[6]{\fnm{Keri} \sur{Hoadley}}\email{khoadley@ufl.edu}
\author[7,12]{\fnm{Min-Young} \sur{Lee}}\email{mlee@kasi.re.kr}
\author[13]{\fnm{Kyoung-Wook} \sur{Min}}\email{kwmin@kaist.ac.kr}
\author[14]{\fnm{Thomas} \sur{M\"{u}ller}}\email{tmueller@mpia.de}
\author[15]{\fnm{Kate} \sur{Pattle}}\email{k.pattle@ucl.ac.uk}
\author[16,17]{\fnm{J. E. G.} \sur{Peek}}\email{jegpeek@stsci.edu}
\author[18]{\fnm{Geoff} \sur{Pleiss}}\email{geoff.pleiss@stat.ubc.ca}
\author[19]{\fnm{David} \sur{Schiminovich}}\email{ds13@columbia.edu}
\author[7,12]{\fnm{Kwang-Il} \sur{Seon}}\email{kiseon@kasi.re.kr}
\author[20]{\fnm{Andrew} \sur{Gordon Wilson}}\email{andrewgw@cims.nyu.edu}
\author[21]{\fnm{Catherine} \sur{Zucker}}\email{catherine.zucker@cfa.harvard.edu}

\affil*[1]{\orgdiv{Department of Physics and Astronomy}, \orgname{Rutgers University}, \orgaddress{\street{136 Frelinghuysen Rd}, \city{Piscataway}, \postcode{08854}, \state{NJ}, \country{USA}}}

\affil[2]{\orgdiv{Center for Computational Astrophysics}, \orgname{Flatiron Institute}, \orgaddress{\street{162 Fifth Avenue}, \city{New York}, \postcode{10010}, \state{NY}, \country{USA}}}

\affil[3]{\orgdiv{Center for Cosmology and Particle Physics}, \orgname{New York University}, \orgaddress{\street{726 Broadway}, \city{New York}, \postcode{10003}, \state{NY}, \country{USA}}}

\affil[4]{\orgdiv{Physics Department}, \orgname{Technion - Israel Institute of Technology}, \orgaddress{\city{Haifa}, \postcode{31000},  \country{Israel}}}

\affil[5]{\orgdiv{Astronomy Unit}, \orgname{Queen Mary University of London}, \orgaddress{\city{London}, \postcode{E14NS},\country{UK}}}

\affil[6]{\orgdiv{Department of Astronomy}, \orgname{University of Florida}, \orgaddress{\street{2-1 Bryant Space Science Center}, \city{Gainesville}, \postcode{32611}, \state{FL}, \country{USA}}}

\affil[7]{\orgdiv{Korea Astronomy \& Space Science Institute},  \orgaddress{\street{776 Daedeok-daero}, \city{Yuseong-gu}, \postcode{34055}, \state{Daejeon}, \country{Republic of Korea}}}

\affil[8]{\orgdiv{McDonald Observatory}, \orgname{University of Texas at Austin}, \orgaddress{\street{2515 Speedway Boulevard}, \city{Austin}, \postcode{78712}, \state{TX}, \country{USA}}}

\affil[9]{\orgdiv{Steward Observatory}, \orgname{University of Arizona}, \orgaddress{\street{933 N Cherry Ave}, \city{Tucson}, \postcode{85719}, \state{AZ}, \country{USA}}}

\affil[10]{\orgdiv{Space Sciences Lab}, \orgname{University of California}, \city{Berkeley}, \postcode{94720}, \state{CA}, \country{USA}}

\affil[11]{\orgdiv{Universit\'{e} Paris Cit\'{e}}, \orgname{Universit\'{e} Paris-Saclay, CEA, CNRS, AIM}, \orgaddress{\city{Gif-sur-Yvette}, \postcode{F-91190}, \country{France}}}

\affil[12]{\orgdiv{Department of Astronomy and Space Science}, \orgname{University of Science and Technology}, \orgaddress{\street{217 Gajeong-ro}, \city{Yuseong-gu}, \postcode{34113}, \state{Daejeon}, \country{Republic of Korea}}}

\affil[13]{ \orgname{ Korea Advanced Institute of Science and Technology (KAIST)}, \orgaddress{\street{291, Daehak-ro, Yuseong-gu} \city{Daejeon}, \postcode{34113}, \country{Repulic of Korea}}}

\affil[14]{ \orgname{ Max Plank Institute for Astronomy (MPIA)}, \orgaddress{\street{ Königstuhl 17} \city{Heidelberg}, \postcode{69117}, \country{Germany}}}

\affil[15]{\orgdiv{Department of Physics and Astronomy}, \orgname{ University College London}, \orgaddress{\street{Gower Street} \city{London}, \postcode{WC1E 6BT}, \country{UK}}}
\affil[16]{\orgdiv{Space Telescope Science Institute},\orgaddress{\city{Baltimore}, \postcode{21218}, \state{MD}, \country{USA}}}

\affil[17]{\orgdiv{Department of Physics \& Astronomy}, \orgname{ Johns Hopkins University}, \orgaddress{\city{Baltimore}, \state{MD} \postcode{21218}, \country{USA}}}

\affil[18]{\orgdiv{University of British Columbia}, \orgaddress{\city{Vancouver}, \state{BC} \postcode{V6T 1Z4}, \country{Canada}}}

\affil[19]{\orgdiv{Department of Astronomy and Columbia Astrophysics Laboratory}, \orgname{Columbia University}, \orgaddress{\street{550 W 120 St. MC 5246-7} \city{New York}, \postcode{10027}, \state{NY}, \country{USA}}}

\affil[20]{\orgdiv{Courant Institute of Mathematical Sciences}, \orgname{New York University}, \orgaddress{\street{251 Mercer St} \city{New York}, \postcode{10012}, \state{NY}, \country{USA}}}

\affil[21]{\orgdiv{Center for Astrophysics $\mid$ Harvard \& Smithsonian}, \orgaddress{\street{60 Garden St.} \city{Cambridge}, \postcode{02138}, \state{MA}, \country{USA}}}


\abstract{
A longstanding prediction in interstellar theory posits that significant quantities of molecular gas, crucial for star formation, may be undetected due to being ``dark" in commonly used molecular gas tracers, such as carbon monoxide.  
We report the discovery of Eos, the closest dark molecular cloud, located just 94 parsecs from the Sun. 
This cloud is the first molecular cloud ever to be identified using H$_2$ far ultra-violet (FUV) fluorescent line emission, which traces molecular gas at the boundary layers of star-forming and supernova remnant regions. 
The cloud edge is outlined along the high-latitude side of the North Polar Spur, a prominent x-ray/radio structure. 
Our distance estimate utilizes 3D dust maps, the absorption of the soft X-ray background, and hot gas tracers such as O\,{\sc vi}; these place the cloud at a distance consistent with the Local Bubble's surface.
Using high-latitude CO maps we note a small amount (M$_{\rm{H}_2}\approx$20-40\,M$_\odot$) of CO-bright cold molecular gas, in contrast with the much larger estimate of the cloud's true molecular mass (M$_{\rm{H}_2}\approx3.4\times 10^3$\,M$_\odot$), indicating most of the cloud is CO-dark. Combining observational data with novel analytical models and simulations, we predict this cloud will photoevaporate in 5.7 million years, placing key constraints on the role of stellar feedback in shaping the closest star-forming regions to the Sun.}




\maketitle

Molecular hydrogen (H$_2$) is the most abundant molecule in the universe and a key ingredient in all known star and planet formation.  The stellar nurseries nearest to the Sun lie along the surface of a structure known as the Local Bubble \cite{2022Natur.601..334Z}, in which our solar system currently resides. Spanning a few hundred parsecs in diameter, the Local Bubble is a superbubble likely formed by multiple supernovae that created a hot evacuated interior cavity surrounded by a shell of swept-up gas and dust  \cite{Berkhuijsen1971,2022Natur.601..334Z}. 



All recent studies of the Local Bubble's star-forming molecular clouds and, indeed, all molecular clouds identified in the Milky Way Galaxy, have relied on 3D and 2D observations of dust in emission and/or extinction \cite{2009A&A...493..735L,Planck2014}, and molecular spectral-line observations of carbon monoxide (CO) \cite{2001ApJ...547..792D,2022ApJS..262....5D,Koda_CO_2023,1982A&A...107..390L,1989ApJ...347..231G} or gamma rays \cite{1982A&A...107..390L,1989ApJ...347..231G}.   CO, in particular, is important because it can be used as a tracer of the total mass of H$_2$ in star-forming clouds, rather than observing the much fainter H$_2$ line emission directly. Observing H$_2$ is challenging because it is a homonuclear molecule and hence has no rotational dipole transitions. The first excited state of H$_2$ capable of emission occurs at T=511K, in stark contrast with the average temperature of T$\approx10$K in dense regions of molecular clouds.  Thus, much of the molecular cloud can not be directly observed in  H$_2$ and other molecules, such as the relatively abundant CO molecule, whose lowest energy level is situated at 5.5K  and OH \cite{Busch_OH_2024}, are often used. Nevertheless, it is possible to directly observe H$_2$ in emission in the infrared (IR) and FUV, where it emits primarily along the boundaries of clouds or in warm shock-excited regions.

Here we identify, for the first time, a new Local Bubble diffuse molecular cloud we name the Eos cloud using the FUV fluorescent emission capabilities of  H$_2$ molecules \textit{directly.} FUV fluorescent emission of  H$_2$ stems from the ability of H$_2$ to absorb FUV photons in the Lyman-Werner band (photons with energy between 11.2eV and 13.6 eV) which sends the molecules into electronic excited states\cite{1987ApJ...322..412B,1989Sternbergb}.   The relaxation back into the electronic ground state gives rise to fluorescent emission lines 
ranging between 912-1700 {\AA}. 

 To reveal the Eos molecular cloud we use the FUV fluorescent survey  Far-ultraviolet Imaging Spectrograph (FIMS). Figure 1 shows the all-sky map of FUV fluorescent H$_2$ emission from FIMS \cite{jo2017ApJS..231...21J}.  FIMS, also known as Spectroscopy of Plasma Evolution from Astrophysical Radiation (SPEAR), hereafter referred to as FIMS/SPEAR, was a far-ultraviolet spectrograph that operated from November 2003 to May 2005 as an instrument on the Korean satellite STSAT-1 \cite{2006ApJ...644L.153E,jo2017ApJS..231...21J}. FIMS/SPEAR observed over 70\% of the sky at moderate spatial (5 arcmin) and low spectral (low spectral resolution (R=$\lambda/\delta\lambda \sim$550). 
 
 In the top two panels, we show the FUV H$_2$ fluorescence emission map, in log line units (LU: photons cm$^{-2}$ sr$^{-1}$ s$^{-1}$). In the bottom panels, we show a map of the ratio of H$_2$ intensity to total FUV intensity, in percentage points (\%). In both the top and bottom panels, the location of the Eos cloud, the feature of interest in this work, is highlighted, and appears as a bright fluorescent feature. The cloud line intensity is  20,000 LU on average and spans  $l = 25^\circ$ -- 45$^\circ$ and $b=40^\circ$ -- 63$^\circ$. The cloud is not present in  H-$\alpha$ emission nor is it seen in the FUV continuum data  (not shown here, but see Figure 3b and Figure 4c from \cite{jo2017ApJS..231...21J}), indicating a lack of massive stars and pointing to its being a gaseous molecular structure. It can therefore be seen very distinctly when viewing the ratio of  H$_2$ fluorescent intensity to total FUV intensity, as seen in the bottom panels of Figure 1.  

 The inserts in Figure 1 show Cartesian close-ups of the cloud.  
 The cloud has a characteristic crescent shape, which can be used to determine its correspondence to other gas tracers. 
For example, the Galactic Arecibo L-Band Feed Array H\,{\sc i} (GALFA-H\,{\sc i}) Survey is a comprehensive project aimed at mapping the neutral hydrogen in the Milky Way at a spatial resolution of 4 arcmins and a spectral resolution of 0.18 $\rm km \; s^{-1}$ \cite{2011ApJS..194...20P}. 
The top right panel of Figure \ref{fig:3ddustmap} shows the contours of the Eos cloud overlaid on the 21-cm GALFA Survey map. 
 H\,{\sc i} is expected to be observed in and around molecular clouds as it is required for shielding molecules from dissociation from background UV photons \cite{2012ApJ...748...75L,2016ApJ...829..102I,2018ApJ...856..136P}. In deeper regions of the cloud, the H$_2$ can self-shield against dissociating UV radiation, but the outer layers will still have substantial amounts of H\,{\sc i} due to ongoing photodissociation \cite{2015ApJ...811L..28B,Sternberg_2021}.  The fluorescent FUV emission traces the atomic-to-molecular cloud boundary (i.e., the H$_2$ to H\,{\sc i} transition), and indeed, an excellent agreement is observed between the 21-cm map and the FIMS/SPEAR contours.  Using a high-latitude CO map \cite{2022ApJS..262....5D}, we can search for CO-emission that might be associated with the Eos cloud. In Figure \ref{fig:3ddustmap}, bottom right, we show the Eos cloud CO data from \cite{2022ApJS..262....5D}.
A small CO cloud is present in the vicinity of the Eos cloud, which is identified as cloud 32 in \cite{2022ApJS..262....5D} with $l=37.75^\circ$ and $b=44.75^\circ$ and is coincident with the location of MBM 40 \cite{Magnani_MBM40_1985, Monaci_MBM40_2023}. We discuss the implications for star formation given the possible association with this CO feature in the Supplemental Methods section.

We compute the distance of the cloud using 3D dust mapping techniques,  estimate its mass and describe its connection to the Local Bubble.  The dust-based total column density maps are shown in Figure 2. Total extinction, derived by integrating the 3D dust mapping technique {\sc Dustribution} \cite{2022A&A...658A.166D, 2023MNRAS.519..228D, 2024MNRAS.tmp.1504D},  is displayed in the top left panel. We further describe the {\sc Dustribution}  algorithm in the Supplemental Methods section. The top middle panel in Figure 2  shows the column density map of the Eos cloud derived from the Planck mission at 545 GHz \cite{2016planck}.  

Using  {\sc Dustribution}, we compute a 3D dust map of the Solar Neighborhood out to a distance of 350~pc. 
Figure 3 shows the reconstructed map, focusing on the region of the Eos cloud
as a function of distance. 
We see a single, distinct cloud that corresponds to the H$_2$ emission stretching from 94 -- 130\,pc; no other clouds are visible along the same lines of sight in Figure 3. This establishes the cloud as among the closest to our Solar System, on the near side of the Sco-Cen OB Association \cite{2022Natur.601..334Z} and consistent with some reported distances to the high-latitude portion of the North Polar Spur known as Loop I from \cite{West_2021_NPS, Panopoulou_2021_NPS}.

Smaller condensations are visible inside the cloud, and it seems that the cloud has a distance gradient, with the parts further from the Galactic plane being closer to us, and vice versa. A projected view of the Eos cloud isosurface and surroundings relative to Local Bubble models is shown in the bottom right of Figure 3. There is no indication of other clouds in the same direction.
Integrating the dust density based on dust properties and a gas-to-dust mass ratio of 124 from \cite{2003ApJ...598.1017D} results in a total dust mass for the cloud of 44\,M$_\odot$ and a total mass of $5.5\times 10^3$\,M$_\odot$ within the magenta contour in Figure 1. If we instead include all the dust in the range $20^{\circ} < l < 50^{\circ}$, $38^{\circ} < b < 70^{\circ}$ and 90 pc $< d <$ 140 pc, we find a total mass of $8.5\times 10^3$\,M$_\odot$. We use \texttt{Astrodendro} to estimate the 3D boundaries of the cloud \cite{2009Natur.457...63G,2013ApJ...770..141B}. From these boundaries, we calculate the effective radius of an equivolume sphere as 25.5~pc. This gives an average mass density of $7.9 \times 10^{-2}$ M$_{\odot}$ pc$^{-3}$. To derive the HI mass of the cloud we convert the HI column density observed within the Eos cloud region assuming an area of $15^{\circ} \times 20^{\circ}$ and $<$N(HI)$> = 3 \times 10^{20}$.  With this we derive a HI mass of $2.0\times 10^3$\,M$_\odot$ taking up $36\%$ of the total cloud mass within the contour.


Interestingly, this molecular cloud is located in the vicinity of a region of the sky with significant emission in radio and X-ray wavelengths known as the North Polar Spur (NPS). The NPS is likely composed of a superposition of emitting layers from as far as the Galactic center to as close as the Local Bubble \cite{Berkhuijsen1971,2022Natur.601..334Z}. The nearby high Galactic latitude portion of the NPS, referred to as Loop I (hereafter NPS/Loop I), is likely at a distance of 105 – 135 pc \cite{West_2021_NPS, Panopoulou_2021_NPS}, although this nearby distance is debated \cite{2014A&A...566A..13P,2023CRPhy..23S...1L}). The 3D dust map from \texttt{Dustribution} as seen in Figure 3 (bottom panels) agrees well with the nearby literature distances showing an arc-like structure corresponding at $\sim 100$ pc. The proximity of both structures to one-another, on-sky and along the line-of-sight may suggest they are directly interacting with one another or that the Eos cloud is a foreground object. In Figures 2 (bottom left and middle panels) and 3 (bottom panels) we show the position of the NPS/Loop I in relation to Eos.
The 0.25 keV soft X-ray map (bottom left panel)  probes hot gas in the local interstellar medium.
The ROSAT 0.25 keV map has been crucial in studying the Local Hot Bubble, showing its extent and structure. Ref. \cite{1991Sci...252.1529S} presented a model for the 0.25 keV soft X-ray diffuse background in which the observed flux is dominated by a $\sim10^6$ K thermal plasma located in a 100 -- 300\,pc diameter bubble surrounding the Sun but has significant contributions from a very patchy Galactic halo and charge exchange in the Heliosphere \cite{Shelton_2008}. 
The Eos cloud's characteristic crescent shape outlines the soft X-ray emission. 
The fact that the Eos cloud shape impacts the soft X-rays detected by ROSAT  may have significant implications for the distance of the cloud and its column density.  The absorption of soft X-rays suggests the nearby nature of the Eos cloud and indicates that the cloud acts as a barrier that prevents the soft X-rays from penetrating it\cite{1991Sci...252.1529S}. 
The bottom middle panel of Figure 2 shows the higher-energy 1 keV map. Interestingly, the Eos cloud contours appear to absorb the 1 keV X-rays of the NPS/Loop I at $b = 40^\circ$ -- $60^\circ$, which is similar to what happens in the 0.25 keV X-ray image.  This may suggest that the cloud is interacting with the X-rays from the NPS/Loop I. 

The shape of the NPS/Loop I emission at these high latitudes is  clearly modulated by the presence of the Eos cloud.  In addition to the ROSAT
soft X-ray data, the Eos cloud can be seen as a potential foreground shadow to other hot gas tracers such as O\,{\sc vi} (see Figure 4). The structural similarities seen in Figure 4 between  the soft X-ray and [O VI] emission, which both intensify at the cloud boundary, are suggestive of an X-ray “halo”.  This is due to the  high ionization potential of O V (113.9 eV, equivalent to $\lambda_{\rm ionization}\approx$ 110\AA\ ) meaning that the generation of O VI ions in the ISM is dominated by collisional ionization and thus requires an interaction of the cloud and hot gas.  While earlier studies \cite{2004ApJ...606..341A}, based on [O VI] absorption towards stars with known distances, have shown that such “X-ray halos” can be physically associated with neutral clouds embedded in hot gas, our [O VI] emission maps are more ambiguous as to the distance to the emitting medium.  The geometric similarities and the need for high energy gas to cause OVI ionization, argues for an association of the cloud with the NPS/Loop I material, whether as solely an X-ray foreground shadow or a true “X-ray halo”.
 In the Methods section, we demonstrate that the fluorescent emission of the cloud is likely generated both by the background ISM FUV radiation field as well as by the X-ray emission coming from the NPS/Loop I. 

\begin{figure*}
    \centering
    \includegraphics[width=\columnwidth]{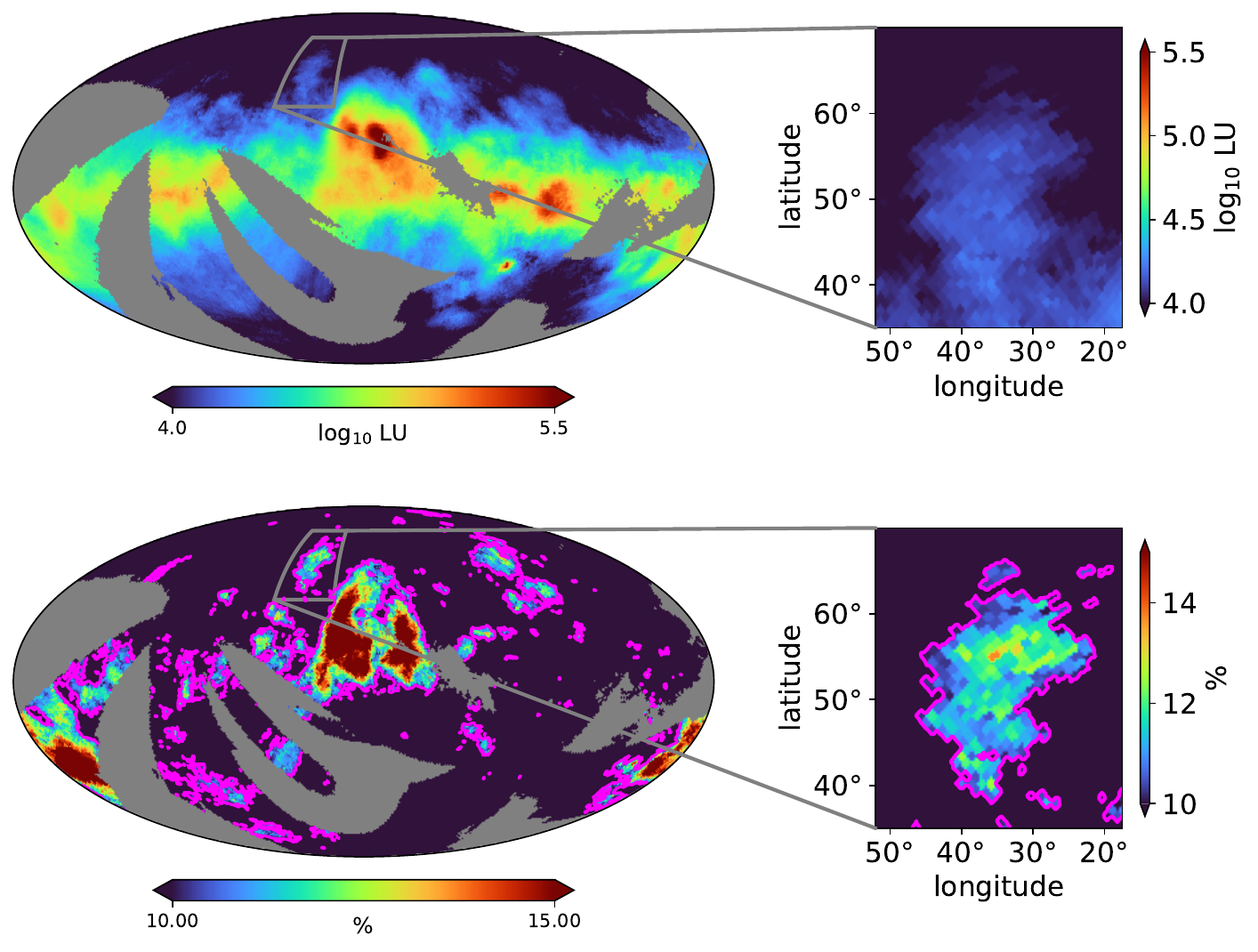}
    \caption{ Top: FIMS/SPEAR FUV H$_2$ fluorescence emission map, in log line units (LU: photons cm$^{-2}$ sr$^{-1}$ s$^{-1}$), first presented by \cite{jo2017ApJS..231...21J}. Bottom: Map of the ratio of H$_2$ intensity to total FUV intensity, in percentage points (\%). In both the top and bottom panels, the location of the Eos cloud is outlined and appears as a bright fluorescent feature that is not present in the FUV continuum. We determine the on-sky boundary for the Eos cloud based on the magenta contours in the bottom panel, which outline the 10\% ratio of H$_{2}$ intensity.}
    \label{fig:joH$_2$map}
\end{figure*}

\begin{figure}
\includegraphics[width=\columnwidth]{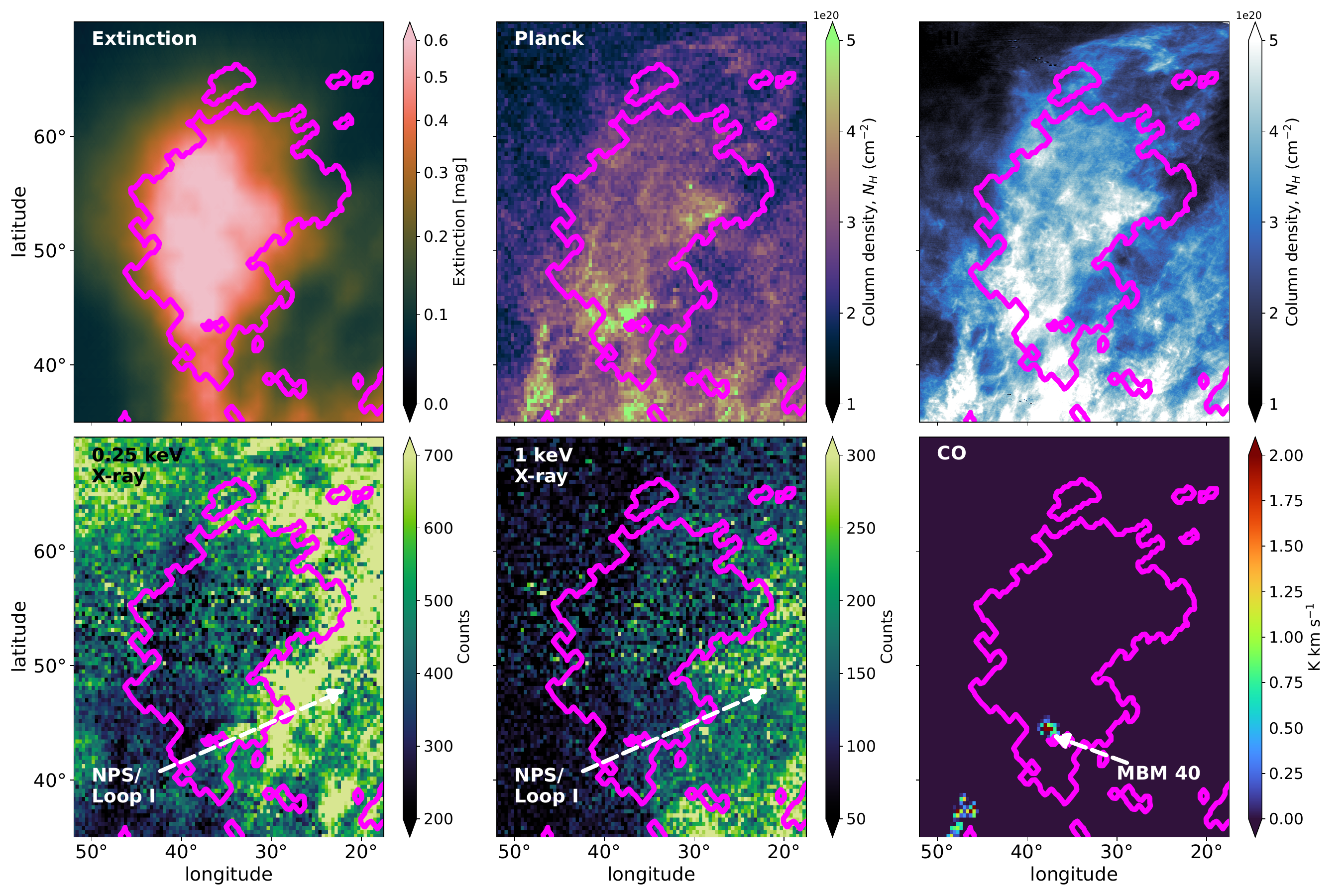}
    \caption{The Eos cloud seen in other tracers. The purple contours in all images represent the H$_2$ emission contour from the fluorescent emission from Fig.\ref{fig:joH$_2$map}. Top Left: Total extinction derived by integrating the {\sc Dustribution} density along the line of sight. Top Middle: Column density map of the Eos cloud derived from Planck 545 GHz data \cite{2016planck}  assuming $T=10$\,K, $\kappa_\nu=4.73$\,cm$^{2}$\,g$^{-1}$ and a distance of 100~pc. Top Right: GALFA-H\,{\sc i} column density map \cite{2011ApJS..194...20P}. Bottom Left and Middle: ROSAT 0.25 keV and 1 keV maps from \cite{1991Sci...252.1529S}. 
      Purple contours are overlaid showing the location of strong H$_2$ fluorescence from Fig.\ref{fig:joH$_2$map}. The Eos cloud shows a prominent outline absorbing the soft X-ray flux and creating a bright X-ray halo towards lower Galactic longitude. The interaction region provides a nearby example of a hot-cold gas interface. Bottom Right: CO data from \cite{2022ApJS..262....5D}; the small CO-bright region (known as MBM 40) within the on-sky cloud boundary is shown by the white arrow. }   \label{fig:3ddustmap}
\end{figure}


\begin{figure}
[h!]
\includegraphics[width=\columnwidth]
{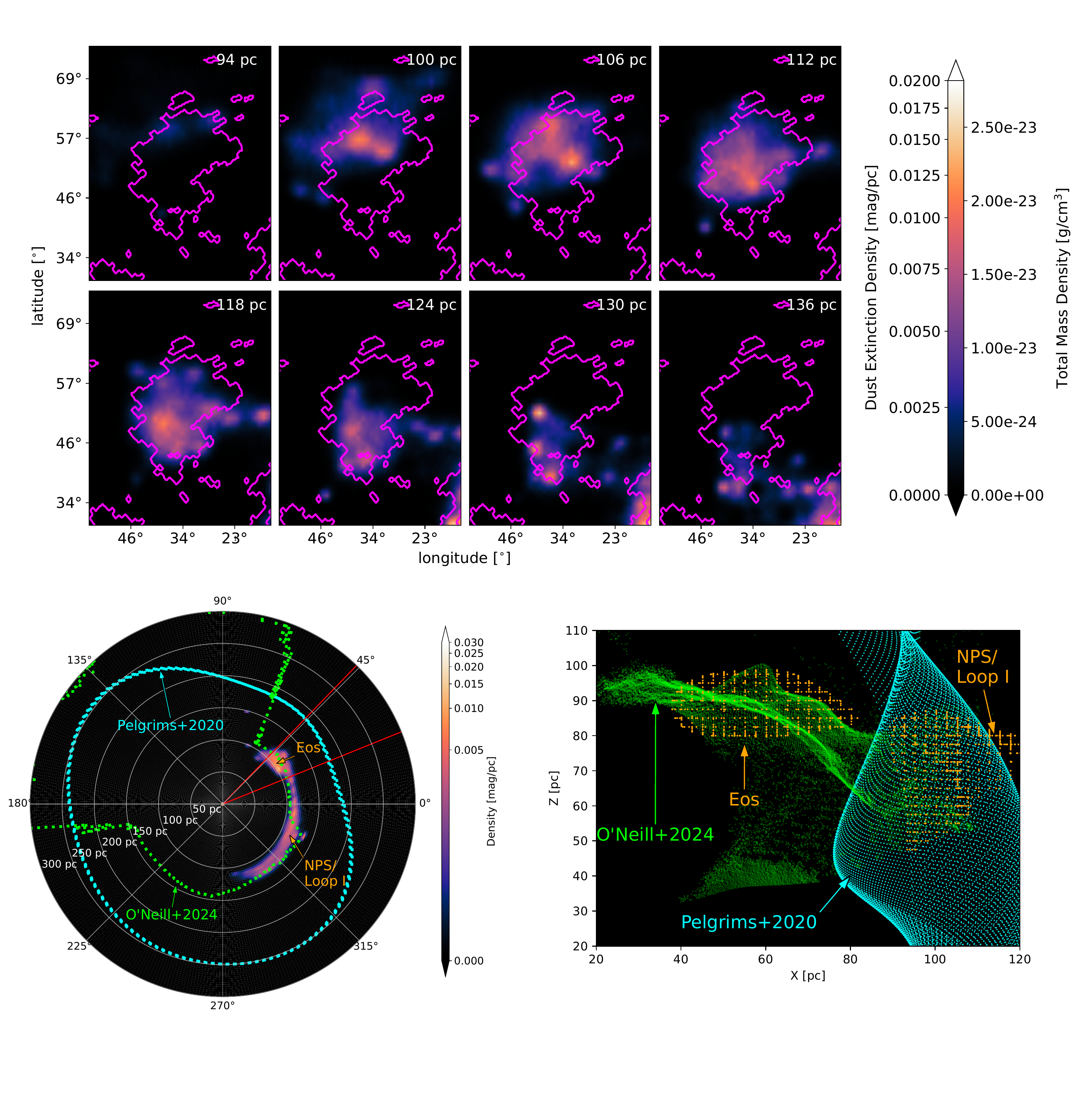}
    \caption{3D dust density structure of the Eos cloud. Top: 2D slices of the cloud at different distances The color bar indicates both the total dust extinction density as well as the total mass density, assuming dust properties from \cite{2003ApJ...598.1017D} and a gas-to-dust mass ratio of 124. As in Figure \ref{fig:joH$_2$map}, the magenta contour shows the region with high H$_2$/FUV ratio. Bottom Left: {\sc Dustribution} density for $b=55^\circ$. The two red lines indicate the minimum and maximum extent of the magenta contour in Fig.~\ref{fig:3ddustmap}, and the cyan and green dashed lines indicate the extent of the Local Bubble at this latitude (from \cite{2020A&A...636A..17P} and \cite{2024arXiv240304961O} respectively). Bottom Right: Projected view of the Eos cloud isosurface and surroundings relative to Local Bubble models. There is no indication of other clouds in the same direction. The 3D dust density of the Solar neighborhood highlighting the Eos cloud and the Local Bubble models can be viewed at links provided in the Data Availability section. }
    \label{fig:dust3D}
\end{figure}



\section*{The Fate of the Eos Cloud}

Will the cloud be photodissociated by the NPS/Loop I X-rays and ISM FUV background photons before star formation can take place?  Regarding the gravitational collapse of the region, we can estimate the thermal Jeans mass of the cloud, assuming a radius of 25.5 pc (estimated using the dust-based distance to the cloud and size on the sky). Masses larger than the Jeans mass are unstable to gravitational contraction, which occurs when the gravitational free-fall timescale is faster than the sound-speed-crossing timescale. Given a range of uncertain temperatures and masses for the cloud, we find that it is marginally stable against gravitational collapse for temperatures typical of warm H$_2$, as described in the Supplemental Methods section. Including the effects of turbulence and magnetic fields in this analysis would only increase stability against collapse.   More precise temperature measurements could be made in the near future with higher spectral resolution FUV instruments, e.g. \cite{2022JATIS...8d4008H}. 

Regarding dissociation, given the average total fluorescent intensity of the Eos cloud and the 
average 21-cm H\,{\sc i} column density, we can compute the dissociation and formation rates of H$_2$ using the framework of \cite{bialy2024}.
Since the process of H$_2$ photo-excitation results in both line emission and H$_2$ dissociation, the H$_2$ dissociation rate is proportional to the total intensity of the H$_2$ emission lines.
We estimate the observed H$_2$ dissociation rate to be $\dot{\Sigma}_{D}^{\rm (obs)}=0.32\ {\rm M_{\odot}  \ pc^{-2} \ Myr^{-1}}$.
Similarly, Equation 12 of \cite{bialy2024} provides an estimate of the formation rate ($\dot{\Sigma}_{F}^{\rm (obs)}$), which we estimate to be
$\dot{\Sigma}_{F}^{\rm (obs)}= 0.02
\ \rm M_{\odot}  \ pc^{-2} \ Myr^{-1}$.

Based on the fact that $\dot{\Sigma}_{F}^{\rm (obs)} < \dot{\Sigma}_{D}^{\rm (obs)}$  and the fact that the cloud is quasi-Jeans supported, we conclude that the cloud is dissociating.    Given $\dot{\Sigma}_{D}^{\rm (obs)}$, a radius of 25 pc, and H$_2$ mass of 3400 M$_\odot$, the cloud will be destroyed on a timescale of $\sim$5.7 Myr and is being photodissociated at a rate of $\sim$600 M$_\odot$/Myr.  

It is of interest to connect these numbers to average conditions at the solar circle. The star formation surface density in the disk local to the Sun is $\sim$0.005 M$_\odot$ yr$^{-1}$ kpc$^2$\cite{Soler2023}.  If the area of disk being surveyed is roughly $200\times 200$ pc$^2$, then the average star formation rate is around $200 M_{\odot}$ Myr$^{-1}$ in this region.    This calculation suggests that the photodissociation rate (and implied gas cycling rate in equilibrium) is about 3 times the star formation rate, which is consistent with typical star formation inefficiencies when combined with dynamical processes such as interstellar turbulence and stellar jets and winds as predicted in the numerical simulations of \cite{federrath2015}. In this respect, dissociating clouds such as the one discovered here may be a common component of the feedback cycle that regulates star formation in galaxies.

\backmatter

\newpage

\bmhead{Methods}
\section*{FIMS/SPEAR Data}

We briefly review the data collection and identification of the H$_2$ lines.  For this investigation, we have used data from the FIMS/SPEAR long-wavelength channel (L-channel; 1350 -- 1710 \AA), which includes several key transitions of molecular hydrogen fluorescence. While the spectral resolution of the data is too low for individual line identification, the collected data provides low-resolution H$_2$ bumps within the L-channel spectrum that can be used to detect H$_2$ fluorescence. The H$_2$ fluorescence features are dominant in two bands, from 1450 -- 1525 \AA~ and from 1560 -- 1630 \AA. Ref. \cite{jo2017ApJS..231...21J} used data from the FIMS/SPEAR mission to construct a nearly all-sky map of diffuse molecular hydrogen fluorescence in emission, as well as an FUV continuum map. 
 
The FIMS/SPEAR all-sky diffuse-background FUV spectrum, weighted by exposure time and with direct stellar photons excluded, consists of multiple components: dust-scattered stellar continuum, hydrogen two-photon continuum, extragalactic background continuum, atomic emission lines, and H\textsubscript{2} fluorescence emission lines. The spectrum includes atomic emission lines such as Si \textsc{iv} $\lambda$1403, Si \textsc{ii}$^*$ $\lambda$1533, C \textsc{iv} $\lambda\lambda$1548, 1551, He \textsc{ii} $\lambda$1640, and Al \textsc{ii} $\lambda$1671, along with several quasi-bandlike features of H\textsubscript{2} fluorescence emission lines. The H\textsubscript{2} fluorescence features are most prominent in the wavelength ranges 1450 -- 1525 \AA\ and 1560 -- 1630 \AA. To improve the signal-to-noise ratio (S/N), the original data cube, with stellar photons removed, was rebinned to the larger wavelength bin size of 3 \AA. As a result, the seven H\textsubscript{2} fluorescence emission lines with significant peaks consist of many narrow lines and appear as broad lines in the coarse-grained spectrum. 
 An example spectrum can be seen in Figure 1 of \cite{jo2017ApJS..231...21J}.  The data is public on the NASA MAST archive.

 To extract only the H\textsubscript{2} fluorescence emission, \cite{jo2017ApJS..231...21J} removed all continuum background components and atomic emission lines. The H\textsubscript{2} fluorescence emission map was constructed using a pixel size of approximately 0.92$^\circ$ (N$_{side}$ = 64, see \cite{gorski1999healpixprimer}). The spectrum for each pixel was obtained by smoothing the spectra of neighboring pixels with weights proportional to the exposure time.
The radius of the smoothing circle for each pixel was adaptively increased from 2$^\circ$ up to 15$^\circ$ in steps of 1$^\circ$ until the S/N per spectral bin was greater than 15.  The Eos cloud is a robust feature of the data regardless of whether adaptive smoothing is utilized or not (see Supplemental Figure 1).

\begin{figure}
    \centering
    \includegraphics[width=\columnwidth, clip=true, trim=0cm 5cm 0cm 5cm]{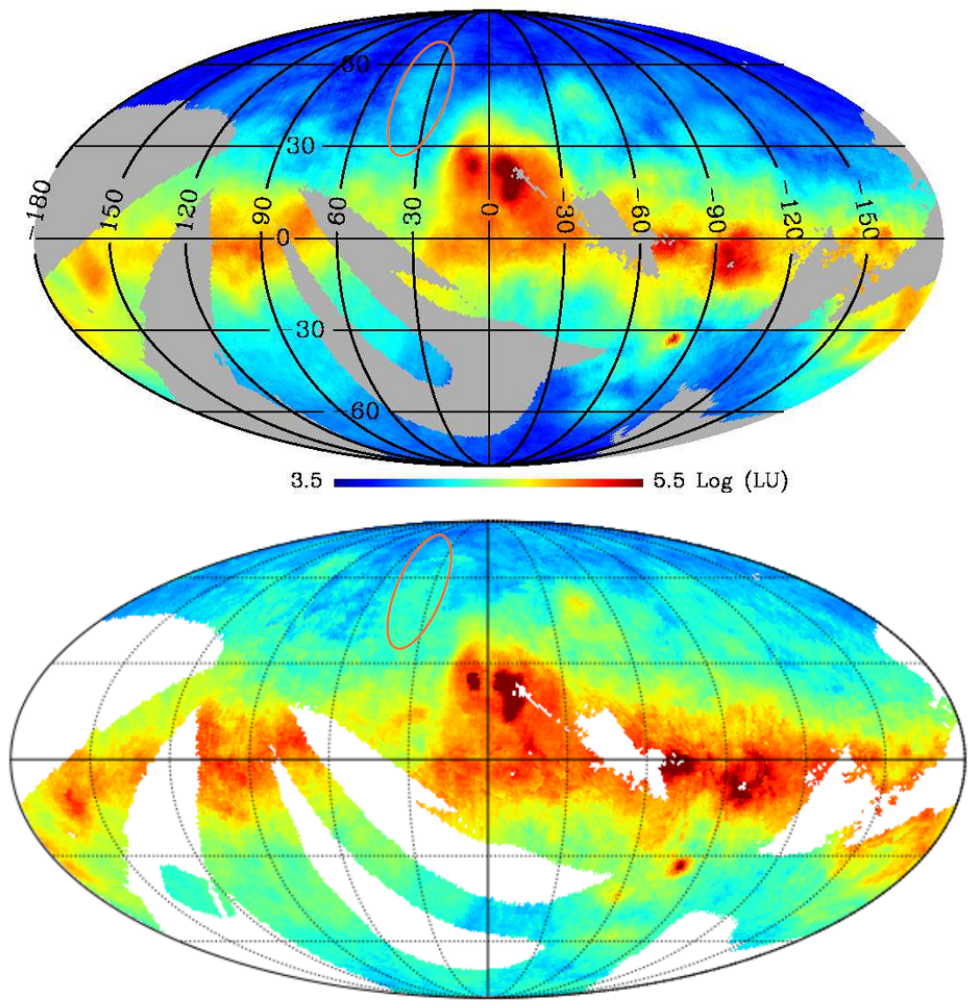}
    \caption{All-sky map of H$_2$ fluorescence with and without the adaptive smoothing method of \cite{jo2017ApJS..231...21J} shown in the top and bottom panels, respectively. The H$_2$ fluorescent emission studied here is a robust feature of the data regardless of whether adaptive smoothing is utilized or not. }
    \label{fig:smoothing}
\end{figure}

\section*{The Origins of the H$_2$ Line Emission}

\subsection*{Modeling the Limited Sensitivity of FIMS/SPEAR}

The Eos Cloud is the first molecular cloud to be discovered through H$_2$ fluorescence lines, which were observed by FIMS/SPEAR during its all-sky survey. A variety of factors limited the instrument's ability to capture all of the H$_2$ fluorescence lines. The primary limitation is the instrument's bandpass, which is unable to capture H$_2$ lines below 1350 \AA. The spectral resolution and sensitivity of the instrument impose further limitations for the remaining in-band emission lines. From the H$_2$ emission map and corresponding exposure time maps developed in \cite{jo2017ApJS..231...21J}, the average exposure time per pixel is $\sim 2200$ seconds for the Eos Cloud. Due to contamination during launch operations, the L-channel sensitivity suffered a loss of $\sim$74\% \cite{2003SPIE.4854..457R,2006ApJ...644L.159E}. Given the relatively low exposure time, significant loss in sensitivity, and low spectral resolution (R$\sim$550), only a small fraction of the total intensity emitted in the H$_2$ lines is detected. We define this fraction to be:

\begin{equation}
\label{eq:eta}
    \eta = \frac{\langle \mathcal{I}_{\rm det} \rangle}{\mathcal{I}_{\rm tot}}
\end{equation}

\noindent where $\mathcal{I}_{\rm tot}$ is the total H$_2$ line intensity emitted by the Eos Cloud and $\langle \mathcal{I}_{\rm det} \rangle$ is the H$_2$ line intensity detected by FIMS/SPEAR, both in LU.    

In order to determine the fraction $\eta$ for the Eos Cloud, we utilize the H2Spec model developed in \cite{2015ApJ...812...41H} to generate synthetic H$_2$ spectra. H2Spec requires as inputs the column density of H$_2$, the gas temperature of H$_2$, and a source spectrum. We assume that the observed H$_2$ is pumped by the Draine UV interstellar radiation field (ISRF) \cite{1978ApJS...36..595D}, which is parameterized by the Draine field strength:

\begin{equation}
\label{eq:chi}
    \chi = \frac{u}{u_0}
\end{equation}

\noindent where $u$ is the FUV energy density within a wavelength range of 912 -- 2480 \AA~ and $u_0$ is the FUV energy density of the Draine ISRF. A value of $\chi=1$ corresponds to a unit Draine field (which is equivalent to G$_0=1.7$, where G$_0$ is the field strength in units of Habing field \cite{1968BAN....19..421H}). We use $\chi = 1$ as the first model input, representative of the typical UV background for the solar neighborhood \cite{1978ApJS...36..595D}, along with an H$_2$ gas temperature of $T=100$ K. We also explored T=500k but found this did not strongly affect the line amplitudes. We generate synthetic fluorescence spectra with H2Spec, where the H$_2$ column density of the emitting layer is the sole variable. The value of $\mathcal{I}_{\rm tot}$ is then calculated for each synthetic spectrum.

To estimate $\langle \mathcal{I}_{\rm det} \rangle$, we model the instrument's response to the synthetic H$_2$ fluorescence spectrum. We first convolve the synthetic spectrum with a line spread function consistent with a fully illuminated instrument slit. We then calculate the noise floor of a single FIMS/SPEAR observation of the Eos Cloud by utilizing the 3$\sigma$ instrument sensitivity curve as a function of wavelength, found in Figure 1 of \cite{2003SPIE.4854..665K}. These sensitivities were calculated before FIMS/SPEAR was launched, and account for the $\sim$74\% loss in sensitivity in the instrument response model. We adjust the sensitivity curve to be consistent with the exposure time reported in the H$_2$ exposure time map. Finally, the observed spectrum is interpolated onto the FIMS/SPEAR L-channel bandpass.

We combine the H2Spec model and the instrument response model to determine the value $\eta$ for the FIMS/SPEAR cloud. We fit the H$_2$ column density to the observed range of $\langle \mathcal{I}_{\rm det} \rangle$ values in the Eos cloud, $\langle \mathcal{I}_{\rm det} \rangle = (1.29 \pm 0.29) \times 10^4$ LU. The best fit 
results in $\mathcal{I}_{\rm tot} = (1.44 \pm 0.10) \times 10^5$ LU. Therefore, the Eos Cloud is observed to have $\eta = (8.96 \pm 2.11) \times 10^{-2}$.
For reference, the value obtained for the model assuming T=500k was $\eta = (7.77 \pm 1.11) \times 10^{-2}$

In the following sections, we compare the value of $\mathcal{I}_{\rm tot}$ with model predictions, first considering an analytic photodissociation region (PDR) model that assumes chemical steady state and then considering an out-of-equilibrium model.

\begin{figure}
    \centering
    \includegraphics[width=0.75\columnwidth]
    {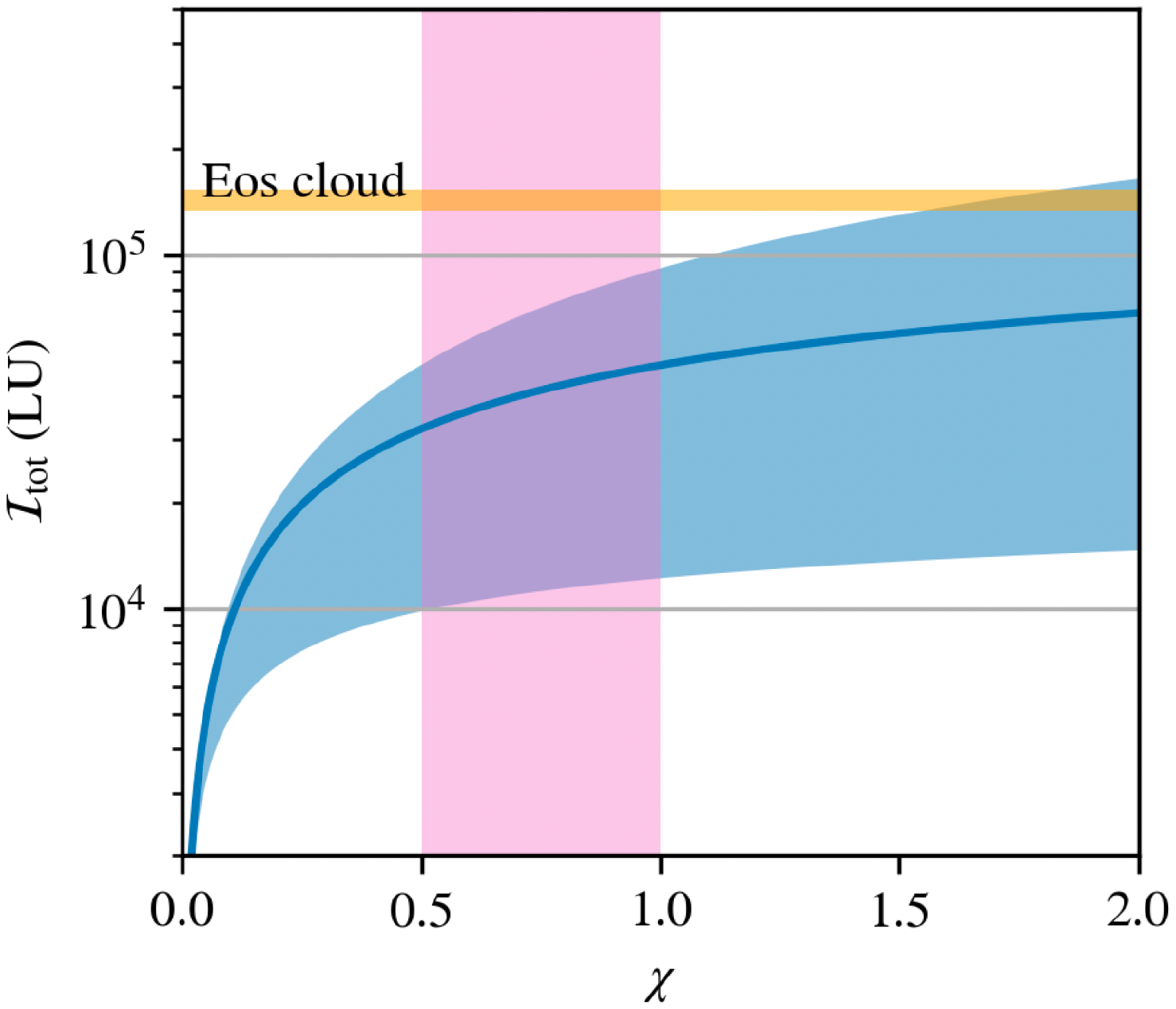}
    \caption{Theoretical predictions of total H$_2$ line intensity ($\mathcal{I}_{\rm tot}$) as a function of the FUV radiation field intensity ($\chi$) in the steady-state PDR model. The blue curve represents $n = 10$ cm$^{-3}$ while the shaded region encompasses densities from $n = 1$ cm$^{-3}$ (lower bound) to $n = 100$ cm$^{-3}$ (upper bound). The vertical magenta band indicates the expected range of $\chi$ values (0.5 to 1) around the standard ISRF. The horizontal orange strip shows the observed total H$_2$ line intensity for the Eos cloud,  
 $\mathcal{I}_{\rm tot} = (1.44 \pm 0.10) \times 10^5$ LU.  Note the general discrepancy between the model predictions and the observed intensity, suggesting additional excitation mechanisms or non-equilibrium conditions in the Eos cloud.}
    
    \label{fig:Itot}
\end{figure}
\subsection*{Chemical Steady-State Theoretical Model}

Following \cite{1989Sternberg}, the total H$_2$ line intensity in a steady-state PDR is given by:
\begin{align}
\label{eq:Itot_1}
    \mathcal{I}_{\rm tot} &= \frac{Rn}{4 \pi \sigma_g}  \frac{1-p_{\rm diss}}{p_{\rm diss}}
    \ln \left( 1+\frac{3 \chi}{n/[10 \ {\rm cm^{-3}}]} \right)  \beta \\ 
    &= 3.5 \times 10^4 \ \left( \frac{n}{10 \ {\rm cm^{-3}}} \right) \ln \left(1+\frac{3\chi}{n/[10 \ {\rm cm^{-3}}]} \right) \ {\rm LU},
\label{eq:Itot_2}
\end{align}
where $\mathcal{I}_{\rm tot}$  (LU) is the total line FUV intensity, $R$ (cm$^3$ s$^{-1}$) is the H$_2$ formation rate coefficient on dust grains, $n$ (cm$^{-3}$) is the hydrogen nucleus number density (including H and H$_2$), $\sigma_g$ (cm$^2$) is the dust absorption cross-section in the Lyman-Werner (LW) band per hydrogen nucleus, $p_{\rm diss}$ is the photodissociation probability per H$_2$ photo-excitation, $\chi$ is the illuminating FUV radiation field intensity in units of the Draine field, and $\beta$ is a dimensionless factor accounting for the attenuation of H$_2$ emission lines by dust (see Appendix A in \cite{bialy2024}).

This model assumes chemical steady state between H$_2$ formation on dust grains and H$_2$ photodissociation by LW radiation within an optically thick, uniform-density 1D slab, externally irradiated by LW radiation normal to the slab surface. For a full derivation, see \cite{1989Sternberg} and Appendix A in \cite{bialy2024}. We express Sternberg's dimensionless parameter $\alpha G$ (characterizing the H$_2$ dissociation-to-formation-rate ratio) in terms of the radiation-intensity-to-density ratio $\chi/n$, assuming $\alpha G = 59 \chi/n \ {\rm cm^3}$ for standard solar metallicity gas (see Eq.~22 in \cite{2016ApJ...822...83B}).

In Eq.~(\ref{eq:Itot_2}), we evaluate the expression assuming standard parameter values: $p_{\rm diss}=0.15$, $R=3\times 10^{-17}$ cm$^3$ s$^{-1}$, $\sigma_g=1.9\times10^{-21}$ cm$^2$, and $\beta=0.5$ \cite{bialy2024}. Notably, $\mathcal{I}_{\rm tot}$ is only weakly sensitive to metallicity and dust-to-gas ratio variations, as both $R$ and $\sigma_g$ are proportional to the effective area of dust grains, causing their effects to largely cancel out in Eq.~(\ref{eq:Itot_2}).

Supplemental Figure 2 presents the theoretical prediction of $\mathcal{I}_{\rm tot}$ as a function of $\chi$ (blue curve) for $n=10$ cm$^{-3}$, as given by Eq.~(\ref{eq:Itot_2}). The surrounding shaded region corresponds to models with varying densities from $n=1$ cm$^{-3}$ (lower envelope) to $n=100$ cm$^{-3}$ (upper envelope). The vertical magenta-shaded zone encloses expected values of $\chi=0.5$ to 1, representing a parameter space around the standard interstellar radiation field (ISRF) value $\chi=1$. The horizontal orange strip above the theoretical models represents the total H$_2$ line intensity emitted by the Eos cloud found in the previous section.

Our analysis reveals that the chemical steady-state model predictions are  below the observed total H$_2$ line intensity emitted by the Eos cloud considering a typical UV ISRF. This discrepancy could be attributed to a combination of factors, including the impact of additional excitation sources, such as energetic photoelectrons produced by X-ray absorption, and the invalidity of the chemical steady-state assumption for the Eos cloud. Both factors are likely contributors, given the cloud's proximity to the NPS/Loop I and its shape imprinted on the NPS/Loop I. 

\subsection*{Non-steady-state Model}


Regarding  out-of-equilibrium H$_2$:
as described in Eq. 9 of \cite{bialy2024}, the observed H$_2$ dissociation rate ($\dot{\Sigma}_{D}^{\rm (obs)}$) is given by
\begin{align}
\label{eq: D-Itot}
    \dot{\Sigma}_{D}^{\rm (obs)} &=  
    0.3 \  \mathcal{I}_5 \left( \frac{N_{21}}{1-\mathrm{e}^{-1.9 N_{21}}} \right)  \ {\rm M_{\odot}  \ pc^{-2} \ Myr^{-1}} \ ,
\end{align}
where $\mathcal{I}_{\rm tot}$ is the total photon intensity summed over all the FUV emission lines (photons cm$^{-2}$ s$^{-1}$ sr$^{-1}$) and 
$\mathcal{I}_{\rm 5} \equiv \mathcal{I}_{\rm tot}/{10^{5} \ ({\rm photons \ cm^{-2} \ s^{-1} \ sr^{-1}}})$, and $N_{21} \equiv N/10^{21} \ ({\rm cm^{-2}})$ is the average column density of atomic neutral hydrogen. Obtaining the estimated averaged $\mathcal{I}_{\rm tot}=1.4\times10^5$ LU from FIMS/SPEAR line modeling and average $N_{21}\sim 0.4$ from GALFA, we get:
$\dot{\Sigma}_{D}^{\rm (obs)}=0.32\ {\rm M_{\odot}  \ pc^{-2} \ Myr^{-1}}$.

Similarly, Eq. 12 of \cite{bialy2024} provides an estimate of the formation rate ($\dot{\Sigma}_{F}^{\rm (obs)}$), with this power law relation:
\begin{align}
\label{eq: JF observation 2}
    \dot{\Sigma}_{F}^{\rm (obs)} &= 0.14 \ f_{\rm H} N_{21}^{1+\alpha} \ {\rm M_{\odot} \ pc^{-2} \ Myr^{-1}} \ ,
\end{align}
where $\alpha=1.3$ and
$f_{\rm H} \equiv N({\rm H})/N$. 
The power law relation with  $N_{21}$ comes from calibration using numerical simulations \cite{2015MNRAS.454..238W,Seifried2017MNRAS.472.4797S} for approximating the product of the formation rate coefficient of H$_2$ on dust grains and the ISM density.  More discussion on the derivation of Eq. \ref{eq: JF observation 2} can be found in \cite{bialy2024}.  For the Eos cloud, the estimated formation rate using $f_{\rm H}=1$ and the average $N_{21} \sim 0.4$ from GALFA is 
$\dot{\Sigma}_{F}^{\rm (obs)}= 0.02
\ \rm M_{\odot}  \ pc^{-2} \ Myr^{-1}$.

These calculations suggest that photodissociation (from X-rays and UV) dominates  H$_2$ formation for this cloud. 
Future studies using higher spectral resolution than that of FIMS/SPEAR could directly test how far the cloud is from chemical equilibrium \cite{2022JATIS...8d4008H}. 

\section*{All-Sky Maps of O\,{\sc vi}, CO and X-Ray Data}

All-sky maps showing the O\,{\sc vi}, CO and X-ray data are provided in Figure 4 to give context to the location of the cloud on the sky in various tracers.  
In particular, in addition to the ROSAT soft X-ray data, the Eos cloud can be seen as a foreground shadow to other hot gas tracers.  In the top panel of Figure 4, we show the cloud contours overlaid on an all-sky map of five-times ionized oxygen (O\,{\sc vi}) published by \cite{2019ApJS..243....9J}. O\,{\sc vi} traces hot, ionized regions probably produced by supernova remnants in the ISM with temperatures around a million degrees
Kelvin. The cloud discussed herein produces a characteristic absorption in the O\,{\sc vi} emission map, similar to what is shown by the soft X-ray data, indicating that it is a cooler, denser foreground object. We note the exact shape of the structures in these emission maps depends on where the hot gas causing the  O\,{\sc vi} ions is coming from and can include geometric effects of the line of sight and magnetic field. O\,{\sc vi} absorption to stars with known distances will be of great value to understanding the relation between the hot gas and the Eos cloud. 
Based on the spatial correspondence between the  X-ray,  O\,{\sc vi} emission map, and the Eos cloud, it's very likely this interaction between the molecular complex and hot gas provides the nearest example of a hot-cold ISM gas interface, which is also thought to be responsible for the O\,{\sc vi} absorption observed through many sight lines throughout the Galaxy.

\begin{figure}
\includegraphics[width=0.75\columnwidth]
{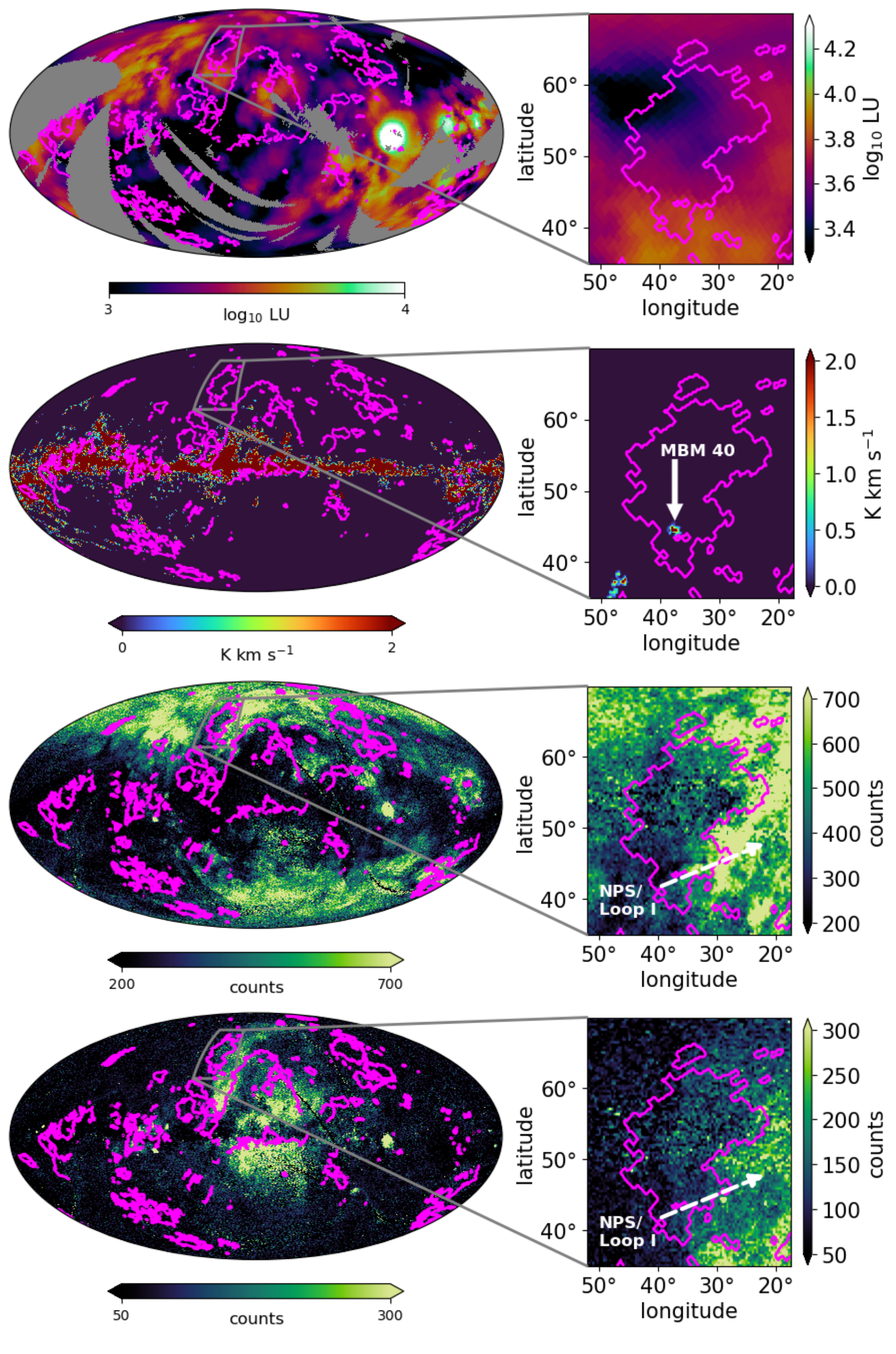}
    \caption{Data from Figure 2, but showing all-sky maps. Top: FIMS/SPEAR O\,{\sc vi} data, modified from \cite{Jo_2019}. The Eos cloud shows a distinctive absorption of hot gas traced by O\,{\sc vi}. The molecular gas blocks hot gas behind it, indicating it is a foreground structure. Upper Middle: CO data from \cite{2022ApJS..262....5D} highlighting the Eos cloud. The small CO-bright region within the on-sky cloud boundary is shown by the white arrow. Lower Middle and Bottom: ROSAT 0.25 keV and 1 keV all-sky maps from \cite{1991Sci...252.1529S}. The Eos cloud shows a prominent outline absorbing the soft X-ray flux and creating a bright X-ray halo towards lower Galactic longitude. The interaction region provides a nearby example of a hot-cold gas interface. Purple contours are overlaid showing the location of strong H$_2$ fluorescence from Fig.\ref{fig:joH$_2$map}. The Eos cloud is outlined and a cut-out zooming in on the region is provided.}
    \label{fig:OVI_CO_softxray}
\end{figure}

\section*{Jeans Stability}

Given the presence of even a small mass of cold CO-bright gas, it is natural to inquire about the fate of this diffuse high-latitude molecular cloud. Is it on its way to being actively star-forming, as is the case of present-day, more massive molecular clouds in the Local Bubble vicinity? Or will the cloud photodissociate before star formation can take place? Using a range of cloud masses and temperatures, we can estimate the Jeans mass of the cloud, assuming a radius of 25.54 pc (estimated using the dust-based distance to the cloud and size on the sky). Masses larger than the Jeans mass are unstable to gravitational contraction, which occurs when the gravitational free-fall timescale is faster than the sound-crossing timescale.

Supplemental Figure 3 shows the ratio of the cloud mass to the Jeans mass, considering only thermal support. For a range of estimated masses from 3D dust maps (Supplemental Figure 4) and over a range of reasonable temperatures for the H$_2$ gas, the cloud is marginally stable against gravitational collapse for temperatures above 100K. This reflects the very low densities of the diffuse gas of  $\rho$ = 0.08 M$_\odot$ pc$^{-3}$.  We note that our calculations make simplistic assumptions about spherical geometry and lack estimations for turbulence and magnetic fields. However, adding these terms would only strengthen the support of the cloud against collapse. Future work will examine the role of turbulence and magnetic fields in the Eos cloud with additional analysis of 21-cm emissions maps and Planck polarization data.

 \begin{figure}
     \centering
     \includegraphics[width=\columnwidth]{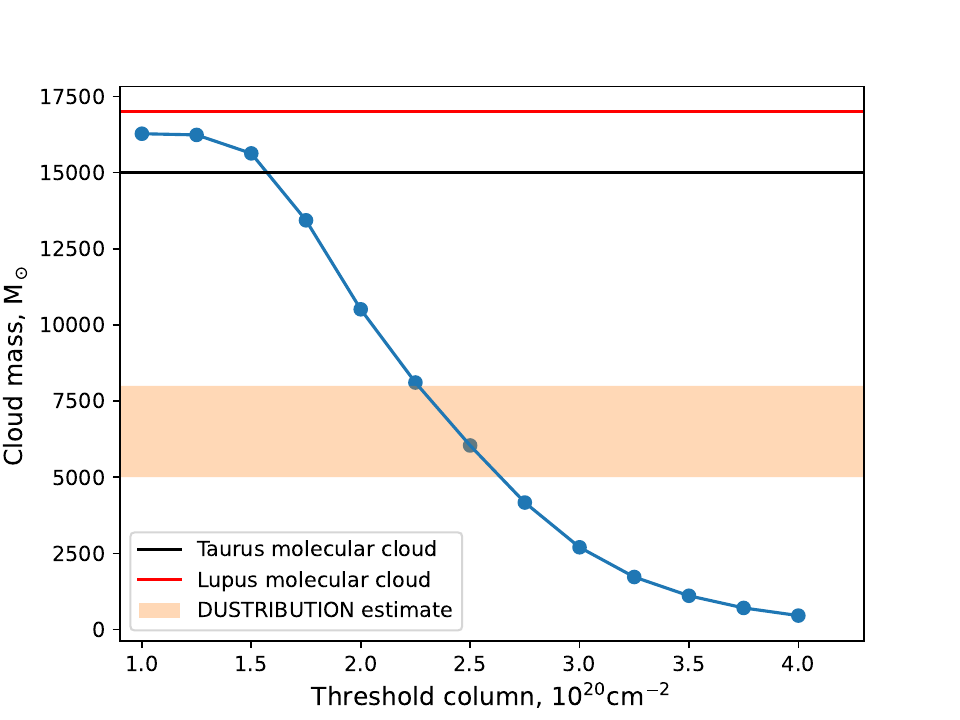}
     \caption{Total mass of the Eos cloud as a function of the threshold column density we use to define the cloud boundary. The cloud extent is best captured by a threshold of $2-3\times10^{20}$\,cm$^{-2}$ and is around half the mass of the Taurus and Lupus molecular clouds. }
     \label{fig:totalMass}
 \end{figure}

\begin{figure}
\includegraphics[width=\columnwidth]
{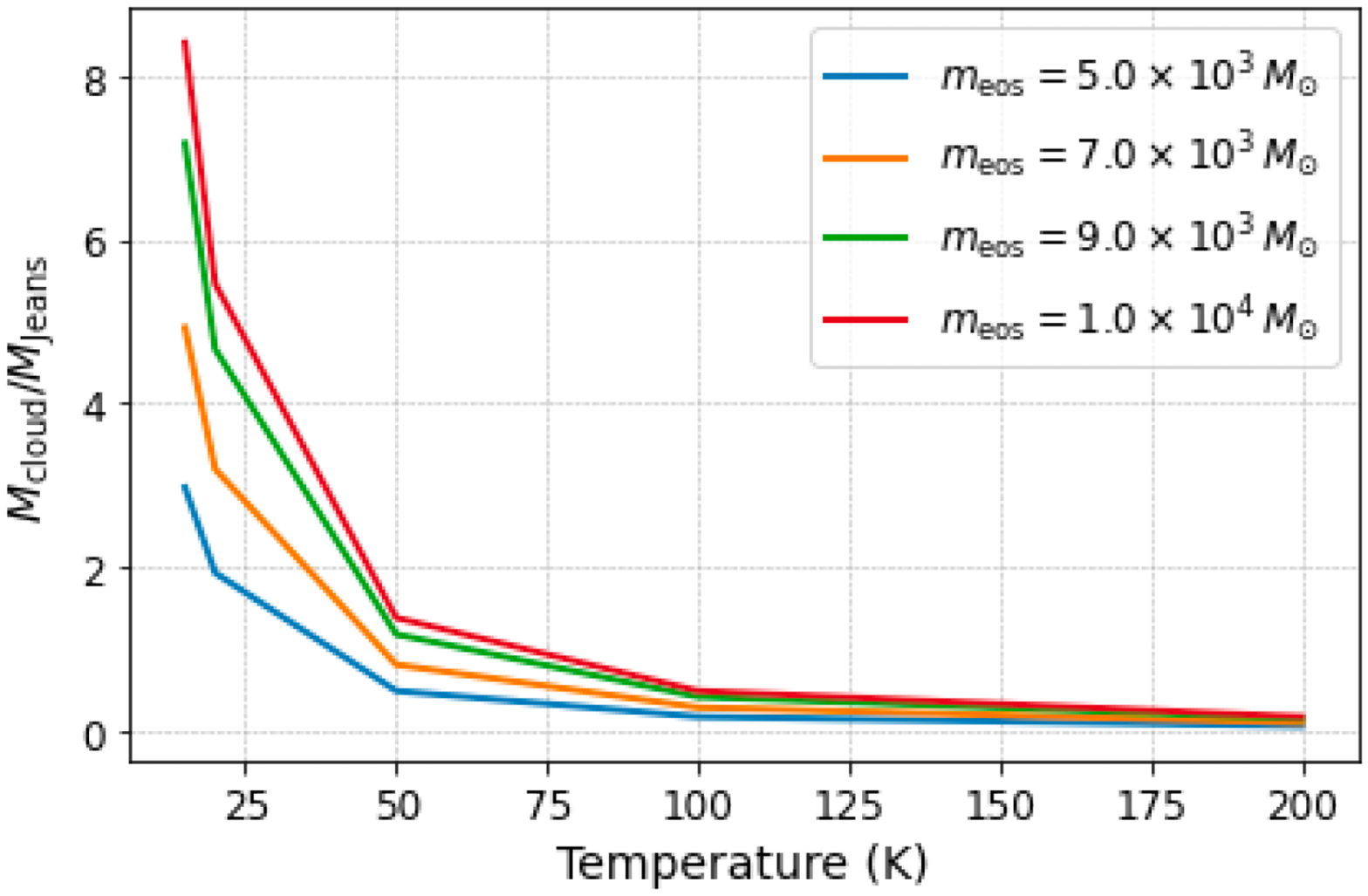}
    \caption{The ratio of cloud mass to Jeans mass ($M_{\mathrm{cloud}}/M_{\mathrm{Jeans}}$) as a function of gas temperature for various cloud masses. The masses estimated from 3D dust are between the blue and orange lines the blue/orange lines. }
    \label{fig:jeans}
\end{figure}

\section*{Building the 3D Dust Density Map}

We compute a 3D dust map of the Solar Neighborhood out to a distance of 350~pc using the 3D dust mapping algorithm {\sc Dustribution} \cite{2022A&A...658A.166D, 2023MNRAS.519..228D, 2024MNRAS.tmp.1504D} modified with a Variational Nearest Neighbor Gaussian process (VNNGP) \cite{wu2022variational}. VNNGPs can scale to almost unlimited numbers of voxels in the map in linear, rather than cubic, time; these modifications will be described in more detail in a forthcoming paper.
This volume is divided into voxels such that $(n_l, n_b, n_d) = (360, 180, 117)$ with equal spacing in $l$ \& $d$ and in $\sin b$,  giving us a grid resolution of $l, b, d = 1^{\circ}, 1^{\circ}, 3$~pc. 

{\sc Dustribution} takes in any catalogue of 3D dust extinction and distances and computes its 3D dust density and extinction within the given region. For the purposes of this paper, we utilize the catalog of \cite{2023MNRAS.524.1855Z}, which derives stellar parameters from Gaia DR 3 BP/RP spectra, BP/RP and G photometry and parallaxes, 2MASS and WISE photometry and LAMOST spectra using a data-driven approach. 

To derive the 3D isocontour from {\sc Astrodendro} requires several input parameters: the minimum density above which cells may be included in structures, the minimum difference at which substructures may be identified, and the minimum number of pixels in a structure, which we set to $4\times10^{-5}$\,mag\,pc$^{-1}$, 0.15\,dex and 8 pixels, respectively, based on the final mean {\sc Dustribution} model parameters. 

For validation, we compare the Eos cloud line-of-sight width and masses to those of \cite{2024A&A...685A..82E}. Applying the above methods to determine distance and mass, we derive a line-of-sight distance of 90-140~pc and a mass of $1.6\times 10^3$\,M$_\odot$. While the line-of-sight width is in excellent agreement with our results from  {\sc Dustribution}, there is a difference of approximately a factor of 3  in the recovered masses. This mass discrepancy could be a result of the differences in the implementation and the resolutions of the two 3D dust density maps.

\section*{CO Mass }
Several large and well-known molecular complexes around the Local Bubble vicinity are actively star-forming, including Taurus, Ophiuchus, Lupus, Chamaeleon, and Corona Australis \cite{2022Natur.601..334Z}.  These clouds are all observed in dense and cold gas tracers such as CO.  In Fig. 4, upper middle, we show the CO data from \cite{2022ApJS..262....5D}.
As previously mentioned, a small CO cloud (MBM 40) is present at $l = 37.75^\circ$ and $b = 44.75^\circ$. 


While the distance to MBM 40 is not well-constrained, if it is associated with the Eos cloud we can compute its mass, given the known distance to Eos.
Adopting a CO-to-H$_2$ conversion factor of 2 $\times 10^{20}$ cm$^{-2}$ (K km s$^{-1}$)$^{-1}$ \cite{2013ARA&A..51..207B}  and including a factor of 1.36 for heavy elements,  we can compute the cloud mass from the CO luminosity:

\begin{equation}
M(M_\odot) = 0.0013 L_{co} \left( \text{K km s}^{-1} \text{deg}^2 \right) d(\text{pc})^2.
\end{equation}
Using $L_{\rm co}=1.53 \ \text{K km s}^{-1} \text{deg}^2$ from \cite{2022ApJS..262....5D} and d = 100~pc, the mass of the Eos cloud predicted by the CO emission is estimated to be $M=19.9 M_\odot$. Variations of a factor of a few in mass may be expected due to uncertainties in the CO-to-H$_2$ conversion factor at high Galactic latitudes \cite{Monaci_MBM40_2023,2024A&A...682A.161S}. The CO luminosity estimates a cloud mass that is 2 -- 3 orders of magnitude off of the mass estimated from dust and traces a much smaller volume than the more diffuse H$_2$ gas. This highlights the importance of tracking CO-dark gas when estimating cloud masses and extents.

\section*{Data Availability}
FIMS/SPEAR data are publicly available to download on the STScI Mikulski Archive for Space Telescopes (MAST) Portal.
The {\sc Dustribution} predicted 3D dust density cube is available to download at \url{www.mwdust.com}. The 3D dust density map of the solar neighborhood including the Eos cloud is available interactively via \url{www.mwdust.com/Eos_Cloud/interactive.html} and a video showing the Eos cloud in relationship to the Sun and local bubble can be found at \url{http://www.mwdust.com/Eos_Cloud/video.mp4}. 

\section*{Code Availability}
The \textsc{Dustribution} code is publicly available in GitHub at \url{www.github.com/Thavisha/Dustribution}. 

\section*{Acknowledgments}
B.B. acknowledges support from NSF grant AST-2407877 and NASA grant No. 80NSSC20K0500. This research was also supported in part by the National Science Foundation under Grant No. NSF PHY-1748958.
B.B. is grateful for generous support from the David and Lucile Packard Foundation, the Alfred P. Sloan Foundation, and the Flatiron Institute, which is funded by the Simons Foundation.
B.B. is very grateful for exciting and enlightening conversations with Christopher McKee, Alex Lazarian, Amiel Sternberg and Bob Benjamin. B.B and T.E.D are very grateful to Julianne Dalcanton for the insightful and exciting discussions. 
T.E.D. acknowledges support for this work provided by NASA through the NASA Hubble Fellowship Program grant No. HST-HF2-51529 awarded by the Space Telescope Science Institute, which is operated by the Association of Universities for Research in Astronomy, Inc., for NASA, under contract NAS 5-26555. 
E.T.H. was supported by internal funding from the University of Arizona, Research Innovation, and Impact.
S.B. acknowledges support from ISF grant 2071540, awarded by the Israel Science Foundation, and the Alon Fellowship (2023), awarded by the Israeli Council for Higher Education.
The authors acknowledge Interstellar Institute's programs and  Paris-Saclay University's Institut Pascal for hosting discussions that nourished the development of the ideas behind this work.
T.J.H. is funded by a Royal Society Dorothy Hodgkin Fellowship and UKRI ERC guarantee funding (EP/Y024710/1).  K.P. is a Royal Society University Research Fellow supported by grant number URF\textbackslash R1\textbackslash 211322.
K.H  acknowledges support through the NASA Roman Technology Fellowship (80NSSC24K0471).
Y.-S.J. was supported by the Basic Science Research Program of the National Research
Foundation of Korea (NRF), funded by the Ministry of Education (2019R1F1A1061102).
K.I.S. was supported by an NRF grant funded by the Korean government (MSIT) (No. 2020R1A2C1005788).
This research is based in part on observations made with the FIMS-SPEAR mission, a joint project of the Korea Advanced Institute of Science and Technology (KAIST), the Korea Astronomy and Space Institute (KASI), and the University of California, Berkeley. FIMS-SPEAR was funded by the Korean Ministry of Science and Technology and NASA grant NAG5-5355. The data reduction for FIMS/SPEAR  was funded by grant NRF-2012M1A3A4A01056418. Data from FIMS-SPEAR were obtained from the MAST data archive at the Space Telescope Science Institute, which is operated by the Association of Universities for Research in Astronomy, Inc., under NASA contract NAS 5–26555.

\section*{Author Contributions}

All correspondences and requests for materials should be addressed to Blakesley Burkhart (bburkhart@flatironinstitute.org).

B.B
Led the overall study, data analysis, and calculations and was the primary author of the manuscript.  

T.D.
Led the analysis of the 3D dust maps and figure creation. Also a primary author of the manuscript editing and writing. 

S.B.
Provided the analysis of the H2-HI equilibrium models and figure. 

T. H. developed the analysis and modeling of the Planck data, including column density/mass estimates and determining whether the cloud is optically thin. He also contributed to the writing of those parts of the text in addition to proof reading of the draft. 

F. A. modeled H2 fluorescence spectra as observed by the FIMS/SPEAR spectrograph. F. Cruz Aguirre additionally wrote the text describing the model and proofread the draft.

Y. J. was the primary producer of the FIMS data used in the paper and FUV emission line maps, including H2 and O VI maps, based on the FIMS/SPEAR mission, and contributed to proofreading the draft and providing scientific comments.

 B. A. provided key insights into the hot/cold interface with OVI. Provided background information about earlier studies and commented on the manuscript.

 H. C.
 contributed to the early phase cloud identification and manuscript editing. 

J. E.
 was the U.S. P.I. of the FIMS/SPEAR mission, designed and managed the optical instrument, payload and mission with the help of the FIMS/SPEAR team.

I. G.
Facilitate comparison of the FUV and X-ray data

E. T. H.
assisted in the initial identification of the Eos cloud from the FIMS/SPEAR data. She also provided notes and discussion on future requirements for observations of the cloud.

W. H.
was the KASI P.I. of the FIMS/SPEAR mission, and contributed to the production of FUV emission line maps, including H2 and O VI maps, based on the FIMS/SPEAR mission.

K. H.
developed H2 synthetic spectral models to predict the UV fluorescence spectrum of ISM H2 populations under various pumping source conditions. Provided guidance and discussion of modeled H2 fluorescence spectra as observed by the FIMS/SPEAR spectrograph, and additionally wrote/provided feedback on the text describing the model and proofread the draft.

M. L.
 produced the HI column density image using the GALFA-HI data and provided scientific comments for the analyses and interpretations.

K. M.
Was the KAIST P.I. of the FIMS/SPEAR mission, and contributed to producing the FUV emission line maps, including the H2 and OVI maps, used in this paper.

T. M.
assisted in 3D dust mapping of the cloud.

K. P.
led analysis of the cloud’s magnetic field properties and analysis of Planck maps.

J. E. G. P.
 contributed to the public release of the FIMS/Spear data on NASA MAST and the discussion on the atomic gas properties of the Eos cloud. 

G. P.
assisted in 3D dust mapping of the cloud.

D. S.
assisted in the initial identification of the Eos cloud from the FIMS/SPEAR data and provided discussion on the fate of the cloud and its connection to the star formation efficiency.

K. S.
was the producer of the FIMS/SPEAR data in the early stage, contributed to the production and release of the FUV emission line maps (H2 and OVI maps), and helped with proofreading the draft and providing scientific comments.

A. G. W.
assisted in 3D dust mapping of the cloud.

C. Z.
contributed to the early phase data analysis discussions of the cloud and proofreading the draft.

\bibliography{ms_eos}

\begin{thebibliography}{10}
\expandafter\ifx\csname url\endcsname\relax
  \def\url#1{\burl{#1}}\fi
\expandafter\ifx\csname urlprefix\endcsname\relax\def\urlprefix{URL }\fi
\providecommand{\bibinfo}[2]{#2}
\providecommand{\eprint}[2][]{\url{#2}}
\providecommand{\doi}[1]{\url{https://doi.org/#1}}
\bibcommenthead

\bibitem{2022Natur.601..334Z}
\bibinfo{author}{{Zucker}, C.} \emph{et~al.}
\newblock \bibinfo{title}{{Star formation near the Sun is driven by expansion of the Local Bubble}}.
\newblock \emph{\bibinfo{journal}{\nat}} \textbf{\bibinfo{volume}{601}}, \bibinfo{pages}{334--337} (\bibinfo{year}{2022}).

\bibitem{Berkhuijsen1971}
\bibinfo{author}{Berkhuijsen, E.~M.}, \bibinfo{author}{Haslam, C. G.~T.} \& \bibinfo{author}{Salter, C.~J.}
\newblock \bibinfo{title}{Structure of the north polar spur and the local supernova hypothesis}.
\newblock \emph{\bibinfo{journal}{Astronomy and Astrophysics}} \textbf{\bibinfo{volume}{14}}, \bibinfo{pages}{252--268} (\bibinfo{year}{1971}).

\bibitem{2009A&A...493..735L}
\bibinfo{author}{{Lombardi}, M.}
\newblock \bibinfo{title}{{NICEST, a near-infrared color excess method tailored to small-scale structures}}.
\newblock \emph{\bibinfo{journal}{\aap}} \textbf{\bibinfo{volume}{493}}, \bibinfo{pages}{735--745} (\bibinfo{year}{2009}).

\bibitem{Planck2014}
\bibinfo{author}{Collaboration, P.}
\newblock \bibinfo{title}{Planck 2013 results. xi. all-sky model of thermal dust emission}.
\newblock \emph{\bibinfo{journal}{\aap}} \textbf{\bibinfo{volume}{571}}, \bibinfo{pages}{A11} (\bibinfo{year}{2014}).
\newblock \urlprefix\url{https://ui.adsabs.harvard.edu/abs/2014A%26A...571A..11P}.

\bibitem{2001ApJ...547..792D}
\bibinfo{author}{{Dame}, T.~M.}, \bibinfo{author}{{Hartmann}, D.} \& \bibinfo{author}{{Thaddeus}, P.}
\newblock \bibinfo{title}{{The Milky Way in Molecular Clouds: A New Complete CO Survey}}.
\newblock \emph{\bibinfo{journal}{\apj}} \textbf{\bibinfo{volume}{547}}, \bibinfo{pages}{792--813} (\bibinfo{year}{2001}).

\bibitem{2022ApJS..262....5D}
\bibinfo{author}{{Dame}, T.~M.} \& \bibinfo{author}{{Thaddeus}, P.}
\newblock \bibinfo{title}{{A CO Survey of the Entire Northern Sky}}.
\newblock \emph{\bibinfo{journal}{\apjs}} \textbf{\bibinfo{volume}{262}}, \bibinfo{pages}{5} (\bibinfo{year}{2022}).

\bibitem{Koda_CO_2023}
\bibinfo{author}{{Koda}, J.}
\newblock \bibinfo{editor}{{Wong}, T.} \& \bibinfo{editor}{{Kim}, W.-T.} (eds) \emph{\bibinfo{title}{{Molecular Clouds with CO-dark Envelopes in the Extended Ultraviolet (XUV) Disk of M83}}}.
\newblock (eds \bibinfo{editor}{{Wong}, T.} \& \bibinfo{editor}{{Kim}, W.-T.}) \emph{\bibinfo{booktitle}{Resolving the Rise and Fall of Star Formation in Galaxies}}, Vol. \bibinfo{volume}{373} of \emph{\bibinfo{series}{IAU Symposium}}, \bibinfo{pages}{15--20} (\bibinfo{year}{2023}).

\bibitem{1982A&A...107..390L}
\bibinfo{author}{{Lebrun}, F.} \emph{et~al.}
\newblock \bibinfo{title}{{COS-B gamma-ray measurements, cosmic rays and the local interstellar medium}}.
\newblock \emph{\bibinfo{journal}{\aap}} \textbf{\bibinfo{volume}{107}}, \bibinfo{pages}{390--396} (\bibinfo{year}{1982}).

\bibitem{1989ApJ...347..231G}
\bibinfo{author}{{Grenier}, I.~A.}, \bibinfo{author}{{Lebrun}, F.}, \bibinfo{author}{{Arnaud}, M.}, \bibinfo{author}{{Dame}, T.~M.} \& \bibinfo{author}{{Thaddeus}, P.}
\newblock \bibinfo{title}{{CO Observations of the Cepheus Flare. I. Molecular Clouds Associated with a Nearby Bubble}}.
\newblock \emph{\bibinfo{journal}{\apj}} \textbf{\bibinfo{volume}{347}}, \bibinfo{pages}{231} (\bibinfo{year}{1989}).

\bibitem{Busch_OH_2024}
\bibinfo{author}{{Busch}, M.~P.}
\newblock \bibinfo{title}{{First Extragalactic Detection of Thermal Hydroxyl (OH) 18 cm Emission in M31 Reveals Abundant CO-faint Molecular Gas}}.
\newblock \emph{\bibinfo{journal}{\apj}} \textbf{\bibinfo{volume}{967}}, \bibinfo{pages}{148} (\bibinfo{year}{2024}).

\bibitem{1987ApJ...322..412B}
\bibinfo{author}{{Black}, J.~H.} \& \bibinfo{author}{{van Dishoeck}, E.~F.}
\newblock \bibinfo{title}{{Fluorescent Excitation of Interstellar H 2}}.
\newblock \emph{\bibinfo{journal}{\apj}} \textbf{\bibinfo{volume}{322}}, \bibinfo{pages}{412} (\bibinfo{year}{1987}).

\bibitem{1989Sternbergb}
\bibinfo{author}{{Sternberg}, A.} \& \bibinfo{author}{{Dalgarno}, A.}
\newblock \bibinfo{title}{{The Infrared Response of Molecular Hydrogen Gas to Ultraviolet Radiation: High-Density Regions}}.
\newblock \emph{\bibinfo{journal}{\apj}} \textbf{\bibinfo{volume}{338}}, \bibinfo{pages}{197} (\bibinfo{year}{1989}).

\bibitem{jo2017ApJS..231...21J}
\bibinfo{author}{{Jo}, Y.-S.}, \bibinfo{author}{{Seon}, K.-I.}, \bibinfo{author}{{Min}, K.-W.}, \bibinfo{author}{{Edelstein}, J.} \& \bibinfo{author}{{Han}, W.}
\newblock \bibinfo{title}{{A Far-ultraviolet Fluorescent Molecular Hydrogen Emission Map of the Milky Way Galaxy}}.
\newblock \emph{\bibinfo{journal}{\apjs}} \textbf{\bibinfo{volume}{231}}, \bibinfo{pages}{21} (\bibinfo{year}{2017}).

\bibitem{2006ApJ...644L.153E}
\bibinfo{author}{{Edelstein}, J.} \emph{et~al.}
\newblock \bibinfo{title}{{The ``Spectroscopy of Plasma Evolution from Astrophysical Radiation'' Mission}}.
\newblock \emph{\bibinfo{journal}{\apjl}} \textbf{\bibinfo{volume}{644}}, \bibinfo{pages}{L153--L158} (\bibinfo{year}{2006}).

\bibitem{2011ApJS..194...20P}
\bibinfo{author}{{Peek}, J.~E.~G.} \emph{et~al.}
\newblock \bibinfo{title}{{The GALFA-HI Survey: Data Release 1}}.
\newblock \emph{\bibinfo{journal}{\apjs}} \textbf{\bibinfo{volume}{194}}, \bibinfo{pages}{20} (\bibinfo{year}{2011}).

\bibitem{2012ApJ...748...75L}
\bibinfo{author}{{Lee}, M.-Y.} \emph{et~al.}
\newblock \bibinfo{title}{{A High-resolution Study of the H I-H$_{2}$ Transition across the Perseus Molecular Cloud}}.
\newblock \emph{\bibinfo{journal}{\apj}} \textbf{\bibinfo{volume}{748}}, \bibinfo{pages}{75} (\bibinfo{year}{2012}).

\bibitem{2016ApJ...829..102I}
\bibinfo{author}{{Imara}, N.} \& \bibinfo{author}{{Burkhart}, B.}
\newblock \bibinfo{title}{{The H I Probability Distribution Function and the Atomic-to-molecular Transition in Molecular Clouds}}.
\newblock \emph{\bibinfo{journal}{\apj}} \textbf{\bibinfo{volume}{829}}, \bibinfo{pages}{102} (\bibinfo{year}{2016}).

\bibitem{2018ApJ...856..136P}
\bibinfo{author}{{Pingel}, N.~M.}, \bibinfo{author}{{Lee}, M.-Y.}, \bibinfo{author}{{Burkhart}, B.} \& \bibinfo{author}{{Stanimirovi{\'c}}, S.}
\newblock \bibinfo{title}{{Multi-phase Turbulence Density Power Spectra in the Perseus Molecular Cloud}}.
\newblock \emph{\bibinfo{journal}{\apj}} \textbf{\bibinfo{volume}{856}}, \bibinfo{pages}{136} (\bibinfo{year}{2018}).

\bibitem{2015ApJ...811L..28B}
\bibinfo{author}{{Burkhart}, B.}, \bibinfo{author}{{Lee}, M.-Y.}, \bibinfo{author}{{Murray}, C.~E.} \& \bibinfo{author}{{Stanimirovi{\'c}}, S.}
\newblock \bibinfo{title}{{The Lognormal Probability Distribution Function of the Perseus Molecular Cloud: A Comparison of HI and Dust}}.
\newblock \emph{\bibinfo{journal}{\apjl}} \textbf{\bibinfo{volume}{811}}, \bibinfo{pages}{L28} (\bibinfo{year}{2015}).

\bibitem{Sternberg_2021}
\bibinfo{author}{Sternberg, A.}, \bibinfo{author}{Gurman, A.} \& \bibinfo{author}{Bialy, S.}
\newblock \bibinfo{title}{H i-to-h transitions in dust-free interstellar gas}.
\newblock \emph{\bibinfo{journal}{The Astrophysical Journal}} \textbf{\bibinfo{volume}{920}}, \bibinfo{pages}{83} (\bibinfo{year}{2021}).
\newblock \urlprefix\url{https://doi.org/10.3847/1538-4357/ac167b}.

\bibitem{Magnani_MBM40_1985}
\bibinfo{author}{{Magnani}, L.}, \bibinfo{author}{{Blitz}, L.} \& \bibinfo{author}{{Mundy}, L.}
\newblock \bibinfo{title}{{Molecular gas at high galactic latitudes.}}
\newblock \emph{\bibinfo{journal}{\apj}} \textbf{\bibinfo{volume}{295}}, \bibinfo{pages}{402--421} (\bibinfo{year}{1985}).

\bibitem{Monaci_MBM40_2023}
\bibinfo{author}{{Monaci}, M.}, \bibinfo{author}{{Magnani}, L.}, \bibinfo{author}{{Shore}, S.~N.}, \bibinfo{author}{{Olofsson}, H.} \& \bibinfo{author}{{Joy}, M.~R.}
\newblock \bibinfo{title}{{Shear, writhe, and filaments: Turbulence in the high-latitude molecular cloud MBM 40}}.
\newblock \emph{\bibinfo{journal}{\aap}} \textbf{\bibinfo{volume}{676}}, \bibinfo{pages}{A138} (\bibinfo{year}{2023}).

\bibitem{2022A&A...658A.166D}
\bibinfo{author}{{Dharmawardena}, T.~E.}, \bibinfo{author}{{Bailer-Jones}, C.~A.~L.}, \bibinfo{author}{{Fouesneau}, M.} \& \bibinfo{author}{{Foreman-Mackey}, D.}
\newblock \bibinfo{title}{{Three-dimensional dust density structure of the Orion, Cygnus X, Taurus, and Perseus star-forming regions}}.
\newblock \emph{\bibinfo{journal}{\aap}} \textbf{\bibinfo{volume}{658}}, \bibinfo{pages}{A166} (\bibinfo{year}{2022}).

\bibitem{2023MNRAS.519..228D}
\bibinfo{author}{{Dharmawardena}, T.~E.} \emph{et~al.}
\newblock \bibinfo{title}{{The three-dimensional structure of galactic molecular cloud complexes out to 2.5 kpc}}.
\newblock \emph{\bibinfo{journal}{\mnras}} \textbf{\bibinfo{volume}{519}}, \bibinfo{pages}{228--247} (\bibinfo{year}{2023}).

\bibitem{2024MNRAS.tmp.1504D}
\bibinfo{author}{{Dharmawardena}, T.~E.} \emph{et~al.}
\newblock \bibinfo{title}{{All-sky three-dimensional dust density and extinction Maps of the Milky Way out to 2.8 kpc}}.
\newblock \emph{\bibinfo{journal}{\mnras}}  (\bibinfo{year}{2024}).

\bibitem{2016planck}
\bibinfo{author}{Adam, R.} \emph{et~al.}
\newblock \bibinfo{title}{Planck2015 results: X. diffuse component separation: Foreground maps}.
\newblock \emph{\bibinfo{journal}{\aap}} \textbf{\bibinfo{volume}{594}}, \bibinfo{pages}{A10} (\bibinfo{year}{2016}).
\newblock \urlprefix\url{http://dx.doi.org/10.1051/0004-6361/201525967}.

\bibitem{West_2021_NPS}
\bibinfo{author}{{West}, J.~L.}, \bibinfo{author}{{Landecker}, T.~L.}, \bibinfo{author}{{Gaensler}, B.~M.}, \bibinfo{author}{{Jaffe}, T.} \& \bibinfo{author}{{Hill}, A.~S.}
\newblock \bibinfo{title}{{A Unified Model for the Fan Region and the North Polar Spur: A Bundle of Filaments in the Local Galaxy}}.
\newblock \emph{\bibinfo{journal}{\apj}} \textbf{\bibinfo{volume}{923}}, \bibinfo{pages}{58} (\bibinfo{year}{2021}).

\bibitem{Panopoulou_2021_NPS}
\bibinfo{author}{{Panopoulou}, G.~V.}, \bibinfo{author}{{Dickinson}, C.}, \bibinfo{author}{{Readhead}, A.~C.~S.}, \bibinfo{author}{{Pearson}, T.~J.} \& \bibinfo{author}{{Peel}, M.~W.}
\newblock \bibinfo{title}{{Revisiting the Distance to Radio Loops I and IV Using Gaia and Radio/Optical Polarization Data}}.
\newblock \emph{\bibinfo{journal}{\apj}} \textbf{\bibinfo{volume}{922}}, \bibinfo{pages}{210} (\bibinfo{year}{2021}).

\bibitem{2003ApJ...598.1017D}
\bibinfo{author}{{Draine}, B.~T.}
\newblock \bibinfo{title}{{Scattering by Interstellar Dust Grains. I. Optical and Ultraviolet}}.
\newblock \emph{\bibinfo{journal}{\apj}} \textbf{\bibinfo{volume}{598}}, \bibinfo{pages}{1017--1025} (\bibinfo{year}{2003}).

\bibitem{2009Natur.457...63G}
\bibinfo{author}{{Goodman}, A.~A.} \emph{et~al.}
\newblock \bibinfo{title}{{A role for self-gravity at multiple length scales in the process of star formation}}.
\newblock \emph{\bibinfo{journal}{\nat}} \textbf{\bibinfo{volume}{457}}, \bibinfo{pages}{63--66} (\bibinfo{year}{2009}).

\bibitem{2013ApJ...770..141B}
\bibinfo{author}{{Burkhart}, B.}, \bibinfo{author}{{Lazarian}, A.}, \bibinfo{author}{{Goodman}, A.} \& \bibinfo{author}{{Rosolowsky}, E.}
\newblock \bibinfo{title}{{Hierarchical Structure of Magnetohydrodynamic Turbulence in Position-position-velocity Space}}.
\newblock \emph{\bibinfo{journal}{\apj}} \textbf{\bibinfo{volume}{770}}, \bibinfo{pages}{141} (\bibinfo{year}{2013}).

\bibitem{2014A&A...566A..13P}
\bibinfo{author}{{Puspitarini}, L.}, \bibinfo{author}{{Lallement}, R.}, \bibinfo{author}{{Vergely}, J.~L.} \& \bibinfo{author}{{Snowden}, S.~L.}
\newblock \bibinfo{title}{{Local ISM 3D distribution and soft X-ray background. Inferences on nearby hot gas and the North Polar Spur}}.
\newblock \emph{\bibinfo{journal}{\aap}} \textbf{\bibinfo{volume}{566}}, \bibinfo{pages}{A13} (\bibinfo{year}{2014}).

\bibitem{2023CRPhy..23S...1L}
\bibinfo{author}{{Lallement}, R.}
\newblock \bibinfo{title}{{North Polar Spur/Loop I: gigantic outskirt of the Northern Fermi bubble or nearby hot gas cavity blown by supernovae?}}
\newblock \emph{\bibinfo{journal}{Comptes Rendus Physique}} \textbf{\bibinfo{volume}{23}}, \bibinfo{pages}{1--24} (\bibinfo{year}{2023}).

\bibitem{1991Sci...252.1529S}
\bibinfo{author}{{Snowden}, S.~L.}, \bibinfo{author}{{Mebold}, U.}, \bibinfo{author}{{Hirth}, W.}, \bibinfo{author}{{Herbstmeier}, U.} \& \bibinfo{author}{{Schmitt}, J.~H.~M.}
\newblock \bibinfo{title}{{ROSAT Detection of an X-ray Shadow in the 1/4-keV Diffuse Background in the Draco Nebula}}.
\newblock \emph{\bibinfo{journal}{Science}} \textbf{\bibinfo{volume}{252}}, \bibinfo{pages}{1529--1532} (\bibinfo{year}{1991}).

\bibitem{Shelton_2008}
\bibinfo{author}{Shelton, R.~L.}
\newblock \bibinfo{title}{Revising the local bubble model due to solar wind charge exchange x-ray emission}.
\newblock \emph{\bibinfo{journal}{Space Science Reviews}} \textbf{\bibinfo{volume}{143}}, \bibinfo{pages}{231–239} (\bibinfo{year}{2008}).
\newblock \urlprefix\url{http://dx.doi.org/10.1007/s11214-008-9358-8}.

\bibitem{2004ApJ...606..341A}
\bibinfo{author}{{Andersson}, B.~G.}, \bibinfo{author}{{Knauth}, D.~C.}, \bibinfo{author}{{Snowden}, S.~L.}, \bibinfo{author}{{Shelton}, R.~L.} \& \bibinfo{author}{{Wannier}, P.~G.}
\newblock \bibinfo{title}{{A Hot Envelope around the Southern Coalsack: X-Ray and Far-Ultraviolet Observations}}.
\newblock \emph{\bibinfo{journal}{\apj}} \textbf{\bibinfo{volume}{606}}, \bibinfo{pages}{341--352} (\bibinfo{year}{2004}).

\bibitem{2020A&A...636A..17P}
\bibinfo{author}{{Pelgrims}, V.}, \bibinfo{author}{{Ferri{\`e}re}, K.}, \bibinfo{author}{{Boulanger}, F.}, \bibinfo{author}{{Lallement}, R.} \& \bibinfo{author}{{Montier}, L.}
\newblock \bibinfo{title}{{Modeling the magnetized Local Bubble from dust data}}.
\newblock \emph{\bibinfo{journal}{\aap}} \textbf{\bibinfo{volume}{636}}, \bibinfo{pages}{A17} (\bibinfo{year}{2020}).

\bibitem{2024arXiv240304961O}
\bibinfo{author}{{O'Neill}, T.~J.}, \bibinfo{author}{{Zucker}, C.}, \bibinfo{author}{{Goodman}, A.~A.} \& \bibinfo{author}{{Edenhofer}, G.}
\newblock \bibinfo{title}{{The Local Bubble is a Local Chimney: A New Model from 3D Dust Mapping}}.
\newblock \emph{\bibinfo{journal}{arXiv e-prints}} \bibinfo{pages}{arXiv:2403.04961} (\bibinfo{year}{2024}).

\bibitem{2022JATIS...8d4008H}
\bibinfo{author}{{Hamden}, E.~T.} \emph{et~al.}
\newblock \bibinfo{title}{{Hyperion: the origin of the stars. A far UV space telescope for high-resolution spectroscopy over wide fields}}.
\newblock \emph{\bibinfo{journal}{Journal of Astronomical Telescopes, Instruments, and Systems}} \textbf{\bibinfo{volume}{8}}, \bibinfo{pages}{044008} (\bibinfo{year}{2022}).

\bibitem{bialy2024}
\bibinfo{author}{Bialy, S.} \emph{et~al.}
\newblock \bibinfo{title}{The molecular cloud lifecycle i: Constraining h2 formation and dissociation rates with observations}.
\newblock \emph{\bibinfo{journal}{\apj}}  (\bibinfo{year}{2024}).
\newblock \urlprefix\url{https://arxiv.org/abs/2408.06416}.

\bibitem{Soler2023}
\bibinfo{author}{{Soler}, J.~D.} \emph{et~al.}
\newblock \bibinfo{title}{{A comparison of the Milky Way's recent star formation revealed by dust thermal emission and high-mass stars}}.
\newblock \emph{\bibinfo{journal}{\aap}} \textbf{\bibinfo{volume}{678}}, \bibinfo{pages}{A95} (\bibinfo{year}{2023}).

\bibitem{federrath2015}
\bibinfo{author}{{Federrath}, C.}
\newblock \bibinfo{title}{{Inefficient star formation through turbulence, magnetic fields and feedback}}.
\newblock \emph{\bibinfo{journal}{\mnras}} \textbf{\bibinfo{volume}{450}}, \bibinfo{pages}{4035--4042} (\bibinfo{year}{2015}).

\bibitem{gorski1999healpixprimer}
\bibinfo{author}{Gorski, K.~M.}, \bibinfo{author}{Wandelt, B.~D.}, \bibinfo{author}{Hansen, F.~K.}, \bibinfo{author}{Hivon, E.} \& \bibinfo{author}{Banday, A.~J.}
\newblock \bibinfo{title}{The healpix primer} (\bibinfo{year}{1999}).
\newblock \urlprefix\url{https://arxiv.org/abs/astro-ph/9905275}.
\newblock \eprint{astro-ph/9905275}.

\bibitem{2003SPIE.4854..457R}
\bibinfo{author}{{Ryu}, K.} \emph{et~al.}
\newblock \bibinfo{editor}{{Blades}, J.~C.} \& \bibinfo{editor}{{Siegmund}, O. H.~W.} (eds) \emph{\bibinfo{title}{{Optics development for the SPEAR mission}}}.
\newblock (eds \bibinfo{editor}{{Blades}, J.~C.} \& \bibinfo{editor}{{Siegmund}, O. H.~W.}) \emph{\bibinfo{booktitle}{Future EUV/UV and Visible Space Astrophysics Missions and Instrumentation.}}, Vol. \bibinfo{volume}{4854} of \emph{\bibinfo{series}{Society of Photo-Optical Instrumentation Engineers (SPIE) Conference Series}}, \bibinfo{pages}{457--466} (\bibinfo{year}{2003}).

\bibitem{2006ApJ...644L.159E}
\bibinfo{author}{{Edelstein}, J.} \emph{et~al.}
\newblock \bibinfo{title}{{The SPEAR Instrument and On-Orbit Performance}}.
\newblock \emph{\bibinfo{journal}{\apjl}} \textbf{\bibinfo{volume}{644}}, \bibinfo{pages}{L159--L162} (\bibinfo{year}{2006}).

\bibitem{2015ApJ...812...41H}
\bibinfo{author}{{Hoadley}, K.}, \bibinfo{author}{{France}, K.}, \bibinfo{author}{{Alexander}, R.~D.}, \bibinfo{author}{{McJunkin}, M.} \& \bibinfo{author}{{Schneider}, P.~C.}
\newblock \bibinfo{title}{{The Evolution of Inner Disk Gas in Transition Disks}}.
\newblock \emph{\bibinfo{journal}{\apj}} \textbf{\bibinfo{volume}{812}}, \bibinfo{pages}{41} (\bibinfo{year}{2015}).

\bibitem{1978ApJS...36..595D}
\bibinfo{author}{{Draine}, B.~T.}
\newblock \bibinfo{title}{{Photoelectric heating of interstellar gas.}}
\newblock \emph{\bibinfo{journal}{\apjs}} \textbf{\bibinfo{volume}{36}}, \bibinfo{pages}{595--619} (\bibinfo{year}{1978}).

\bibitem{1968BAN....19..421H}
\bibinfo{author}{{Habing}, H.~J.}
\newblock \bibinfo{title}{{The interstellar radiation density between 912 A and 2400 A}}.
\newblock \emph{\bibinfo{journal}{\bain}} \textbf{\bibinfo{volume}{19}}, \bibinfo{pages}{421} (\bibinfo{year}{1968}).

\bibitem{2003SPIE.4854..665K}
\bibinfo{author}{{Korpela}, E.~J.} \emph{et~al.}
\newblock \bibinfo{editor}{{Blades}, J.~C.} \& \bibinfo{editor}{{Siegmund}, O. H.~W.} (eds) \emph{\bibinfo{title}{{The SPEAR science payload}}}.
\newblock (eds \bibinfo{editor}{{Blades}, J.~C.} \& \bibinfo{editor}{{Siegmund}, O. H.~W.}) \emph{\bibinfo{booktitle}{Future EUV/UV and Visible Space Astrophysics Missions and Instrumentation.}}, Vol. \bibinfo{volume}{4854} of \emph{\bibinfo{series}{Society of Photo-Optical Instrumentation Engineers (SPIE) Conference Series}}, \bibinfo{pages}{665--675} (\bibinfo{year}{2003}).

\bibitem{1989Sternberg}
\bibinfo{author}{{Sternberg}, A.}
\newblock \bibinfo{title}{{Ultraviolet Fluorescent Molecular Hydrogen Emission}}.
\newblock \emph{\bibinfo{journal}{\apj}} \textbf{\bibinfo{volume}{347}}, \bibinfo{pages}{863} (\bibinfo{year}{1989}).

\bibitem{2016ApJ...822...83B}
\bibinfo{author}{{Bialy}, S.} \& \bibinfo{author}{{Sternberg}, A.}
\newblock \bibinfo{title}{{Analytic H I-to-H$_{2}$ Photodissociation Transition Profiles}}.
\newblock \emph{\bibinfo{journal}{\apj}} \textbf{\bibinfo{volume}{822}}, \bibinfo{pages}{83} (\bibinfo{year}{2016}).

\bibitem{2015MNRAS.454..238W}
\bibinfo{author}{{Walch}, S.} \emph{et~al.}
\newblock \bibinfo{title}{{The SILCC (SImulating the LifeCycle of molecular Clouds) project - I. Chemical evolution of the supernova-driven ISM}}.
\newblock \emph{\bibinfo{journal}{\mnras}} \textbf{\bibinfo{volume}{454}}, \bibinfo{pages}{238--268} (\bibinfo{year}{2015}).

\bibitem{Seifried2017MNRAS.472.4797S}
\bibinfo{author}{{Seifried}, D.} \emph{et~al.}
\newblock \bibinfo{title}{{SILCC-Zoom: the dynamic and chemical evolution of molecular clouds}}.
\newblock \emph{\bibinfo{journal}{\mnras}} \textbf{\bibinfo{volume}{472}}, \bibinfo{pages}{4797--4818} (\bibinfo{year}{2017}).

\bibitem{2019ApJS..243....9J}
\bibinfo{author}{{Jo}, Y.-S.} \emph{et~al.}
\newblock \bibinfo{title}{{Global Distribution of Far-ultraviolet Emissions from Highly Ionized Gas in the Milky Way}}.
\newblock \emph{\bibinfo{journal}{\apjs}} \textbf{\bibinfo{volume}{243}}, \bibinfo{pages}{9} (\bibinfo{year}{2019}).

\bibitem{Jo_2019}
\bibinfo{author}{Jo, Y.-S.} \emph{et~al.}
\newblock \bibinfo{title}{Global distribution of far-ultraviolet emissions from highly ionized gas in the milky way}.
\newblock \emph{\bibinfo{journal}{The Astrophysical Journal Supplement Series}} \textbf{\bibinfo{volume}{243}}, \bibinfo{pages}{9} (\bibinfo{year}{2019}).
\newblock \urlprefix\url{http://dx.doi.org/10.3847/1538-4365/ab22ae}.

\bibitem{wu2022variational}
\bibinfo{author}{Wu, L.}, \bibinfo{author}{Pleiss, G.} \& \bibinfo{author}{Cunningham, J.~P.}
\newblock \bibinfo{title}{Variational nearest neighbor gaussian process}.
\newblock \emph{\bibinfo{journal}{International Conference on Machine Learning}} \bibinfo{pages}{24114--24130} (\bibinfo{year}{2022}).

\bibitem{2023MNRAS.524.1855Z}
\bibinfo{author}{{Zhang}, X.}, \bibinfo{author}{{Green}, G.~M.} \& \bibinfo{author}{{Rix}, H.-W.}
\newblock \bibinfo{title}{{Parameters of 220 million stars from Gaia BP/RP spectra}}.
\newblock \emph{\bibinfo{journal}{\mnras}} \textbf{\bibinfo{volume}{524}}, \bibinfo{pages}{1855--1884} (\bibinfo{year}{2023}).

\bibitem{2024A&A...685A..82E}
\bibinfo{author}{{Edenhofer}, G.} \emph{et~al.}
\newblock \bibinfo{title}{{A parsec-scale Galactic 3D dust map out to 1.25 kpc from the Sun}}.
\newblock \emph{\bibinfo{journal}{\aap}} \textbf{\bibinfo{volume}{685}}, \bibinfo{pages}{A82} (\bibinfo{year}{2024}).

\bibitem{2013ARA&A..51..207B}
\bibinfo{author}{{Bolatto}, A.~D.}, \bibinfo{author}{{Wolfire}, M.} \& \bibinfo{author}{{Leroy}, A.~K.}
\newblock \bibinfo{title}{{The CO-to-H$_{2}$ Conversion Factor}}.
\newblock \emph{\bibinfo{journal}{\araa}} \textbf{\bibinfo{volume}{51}}, \bibinfo{pages}{207--268} (\bibinfo{year}{2013}).

\bibitem{2024A&A...682A.161S}
\bibinfo{author}{{Skalidis}, R.}, \bibinfo{author}{{Goldsmith}, P.~F.}, \bibinfo{author}{{Hopkins}, P.~F.} \& \bibinfo{author}{{Ponnada}, S.~B.}
\newblock \bibinfo{title}{{Constraining the H$_{2}$ column densities in the diffuse interstellar medium using dust extinction and H I data}}.
\newblock \emph{\bibinfo{journal}{\aap}} \textbf{\bibinfo{volume}{682}}, \bibinfo{pages}{A161} (\bibinfo{year}{2024}).

\end{thebibliography}


\end{document}